\documentclass[twoside]{article}

\usepackage[USenglish]{babel}
\usepackage{titling}
\usepackage{authblk}
\usepackage{verbatim}
\usepackage{listings}
\usepackage[utf8]{inputenc}
\usepackage[T1]{fontenc}
\usepackage{epstopdf}
\usepackage{graphicx,graphics}
\usepackage{float}
\usepackage{caption}
\usepackage{amsmath,amsfonts,amssymb}
\usepackage[table,dvipsnames]{xcolor}
\usepackage{geometry}
\usepackage{hyperref}
\usepackage{textcomp}
\usepackage{diagbox}
\usepackage{tabularx}
\usepackage{forest}
\usepackage{parskip}
\usepackage{pgf-umlsd}
\usepackage{pdfpages}
\usepackage{pdflscape}
\usepackage{capt-of}
\usepackage{soul}
\usepackage{amsbsy,marvosym,threeparttable,amsthm,subfigure}
\usepackage{eurosym,mathrsfs,fancyhdr,multicol,indentfirst,color,bm,upgreek,booktabs}
\usepackage{picins}

\renewcommand{\thefootnote}{\fnsymbol{footnote}}
\def\footnoterule{\kern 1mm \hrule width 10cm \kern 2mm}

\allowdisplaybreaks
\catcode`@=11
\def\title#1{\vspace{3mm}\begin{flushleft}\vglue-.1cm\Large\bf\boldmath\protect\baselineskip=18pt plus.2pt minus.1pt #1 \end{flushleft}\vspace{1mm} }
 \def\author#1{\begin{flushleft}\normalsize #1\end{flushleft}\vspace*{-4pt} \vspace{3mm}}
\def\address#1#2{\begin{flushleft}\vglue-.35cm${}^{#1}$\small\it #2\vglue-.35cm\end{flushleft}\vspace{-2mm}\par}

\catcode`@=11
 \def\section{\@startsection{section}{1}{\z@}%
  {-3ex \@plus -.3ex \@minus -.2ex}%
  {2.2ex \@plus.2ex}%
 {\normalfont\normalsize\protect\baselineskip=14.5pt plus.2pt minus.2pt\bfseries}}
 \def\subsection{\@startsection{subsection}{2}{\z@}%
  {-3ex\@plus -.2ex \@minus -.2ex}%
  {2ex \@plus.2ex}%
 {\normalfont\normalsize\protect\baselineskip=12.5pt plus.2pt minus.2pt\bfseries}}
 \def\subsubsection{\@startsection{subsubsection}{3}{\z@}%
  {-2.2ex\@plus -.21ex \@minus -.2ex}%
  {1.4ex \@plus.2ex}
 {\normalfont\normalsize\protect\baselineskip=12pt plus.2pt minus.2pt\sl}}
 
\usetikzlibrary{arrows}
\usetikzlibrary{positioning}

\makeatletter
\renewcommand{\thesection}{\@arabic\c@section}
\newcommand\tinysize{\@setfontsize\tinysize\@vipt\@viipt}
\makeatother

\tikzset{
  my blue box/.style={fill=gray!20, rectangle, rounded corners=3pt},
}

\forestset{
  my blue label/.style={
    label={[my blue box]left:#1},
    s sep+=10pt,
  }
}

\begin{document}
\setstcolor{red}
\thispagestyle{empty}

\title{Personal information self-management: A survey of technologies supporting administrative services}\vspace*{3mm}

\author{Paul Marillonnet$^{1,2,*}$, Maryline Laurent$^{2}$ and Mikaël Ates$^{1}$}\vspace*{1mm}

\address{1}{Entr'ouvert, Paris 75014, France}\vspace*{4mm}
\address{2}{SAMOVAR, Télécom SudParis, Institut Polytechnique de Paris, Évry 91000, France}\vspace*{1mm}

\noindent E-mail: pmarillonnet@entrouvert.com; maryline.laurent@telecom-sudparis.eu; mates@entrouvert.com \\[-1mm]

\thispagestyle{empty}
\let\thefootnote\relax\footnotetext{{}\\[-4mm]\indent\ Regular Paper\\[.5mm]
\indent\ $^*$Corresponding Author}

\newcommand{\thefootnote}{\arabic{footnote}}
\noindent {\small\bf Abstract} \quad {\small
This paper presents a survey of technologies for personal data self-management interfacing with administrative and territorial public service providers.
It classifies a selection of scientific technologies into four categories of solutions: Personal Data Store (PDS), Identity Manager (IdM), Anonymous Certificate System and Access Control Delegation Architecture.
Each category, along with its technological approach, is analyzed thanks to eighteen identified functional criteria that encompass architectural and communication aspects, as well as user data lifecycle considerations.\\
The originality of the survey is multifold.
First, as far as we know, there is no such thorough survey covering such a panel of a dozen of existing solutions.
Second, it is the first survey addressing Personally Identifiable Information (PII) management for both administrative and private service providers.
Third, this paper achieves a functional comparison of solutions of very different technical natures.\\
The outcome of this paper is the clear identification of functional gaps of each solution.
As a result, this paper establishes the research directions to follow in order to fill these functional gaps.
}

\vspace*{3mm}


\vspace*{4mm}

\baselineskip=18pt plus.2pt minus.2pt
\parskip=0pt plus.2pt minus0.2pt
\begin{multicols}{2}

\section{Introduction\label{intro}}
Recent events of privacy infringement for Web users including Equifax~\footnote{See \textit{As Equifax Amassed Ever More Data, Safety Was a Sales Pitch}: \url{https://www.nytimes.com/2017/09/23/business/equifax-data-breach.html} (The New York Times; last accessed: March 10, 2021).},
Cambridge Analytica~\footnote{See \textit{50 million Facebook profiles harvested for Cambridge Analytica in major data breach}: \url{https://www.theguardian.com/news/2018/mar/17/cambridge-analytica-facebook-influence-us-election} (The Guardian; last accessed: March 10, 2021).}
and hacking of Web giants~\footnote{
See \textit{Yahoo says all three billion accounts hacked in 2013 data theft}: \url{https://www.reuters.com/article/us-yahoo-cyber/yahoo-says-all-three-billion-accounts-hacked-in-2013-data-theft-idUSKCN1C82O1} (Reuters; last accessed: March 10, 2021)
and \textit{eBay faces investigations over massive data breach}: \url{http://www.bbc.com/news/technology-27539799} (BBC; last accessed: March 10, 2021).}
reveal the necessity for empowering users with their personal data governance, by letting them manage their Personally Identifiable Information (PII) along their full lifecycle.

The PII leaked during these privacy infringement events can be bank information or any other kind of information that has a monetary value, either directly or indirectly (\textit{e.g.}, through user impersonation fraud) -- \textit{i.e.}, users' addresses, identity documents, phone line documents or miscellaneous contractual documents.

This need for digital self-determination relative to personal data has been consolidated by the General Data Protection Regulation (GDPR)\footnote{See the \textit{Regulation (EU) 2016/679 of the European Parliament and of the Council of 27 April 2016 on the protection of natural persons with regard to the processing of personal data and on the free movement of such data, and repealing Directive 95/46/EC (General Data Protection Regulation)}: \url{https://eur-lex.europa.eu/legal-content/EN/ALL/?uri=CELEX\%3A32016R0679} (last accessed: March 10, 2021).}, enforced on May 2018.

This paper proposes a survey on technologies for user self-management of PII in the context of online administrative services.
These services are  fed with true sensitive PII (name, address, family status, allowances, tax status, {\it etc.}).
Moreover, services are interacting with other private service providers, as they are used to ensure CRM with services offered by third-parties.

The paper is organized as follows.
Section \ref{fundamentals} describes the system model along with the main use case, illustrating the need for PII self-management.
Section \ref{criteres} identifies the eighteen differentiating criteria selected for this comparative survey.
Section \ref{taxonomy} introduces the four categories of solutions selected for the survey.
The main technical background for each solution is also described.
The solutions share the common objective of providing Web users with some informational governance tools, but they differ by the scope of supported functional mechanisms.
These solutions stand out by the way they deal with PII management -- each one offering its own functional mechanisms.
Section \ref{evaluation} evaluates each of the four categories of solutions against the eighteen criteria.
Section \ref{synthese} gives a full synthesis of the survey.
This section also identifies the functions that remain unsupported so far, as well as describing an optimal solution for the main use case.
Before concluding in Section \ref{conclusions}, Section \ref{research-directions} defines the research directions in order to fill the functional gaps identified in this survey.

\paragraph{Acronyms:}
Please refer to Table~\ref{acronymtab} for a complete list of the acronyms used in this article.
\begin{table}[H]
{\scriptsize
\caption{List of Acronyms}
\begin{tabular}{ll}
\hline\hline\hline
{\bf Acronym} & {\bf Definition} \\
\hline
AC & Access Control \\
ACL & Access-Control List \\
API & Application Programming Interface \\
ASP & Administrative Service Provider \\
C & Client \\
CRL & Certificate-Revocation List \\
CRM & Citizen-Relationship Management \\
CRUD & Create, Read, Update, Delete \\
FIM & Federated Identity Management \\
GDPR & General Data Protection Regulation \\
GUI & Graphical User Interface \\
IDaaS & Identity as a Service \\
IdM & Identity Manager \\
JSON & JavaScript Object Notation \\
PDS & Personal Data Store \\
PII & Personally Identifiable Information \\
PSP & Private Service Provider \\
RBAC & Role-Based Access-Control \\
REST & Representational State Transfer \\
SaaS & Software as a Service \\
SAML & Security Assertion Markup Language \\
SLO & Single Logout \\
SoC & System on Chip \\
SP & Service Provider \\
SSO & Single Sign On \\
TTL & Time To Live \\
TTS & Token Translation System \\
UI & User Interface \\
UMA & User-Managed Access \\
VPN & Virtual Private Network \\
XML & eXtensible Markup Language \\
ZKP & Zero-Knowledge Proofs \\
\hline\hline\hline
\end{tabular}
\label{acronymtab}}
\end{table}

\section{Fundamentals: system model and use case\label{fundamentals}}

\subsection{System and security models\label{models}}
\begin{figure*}[!htb]
\begin{center}
\includegraphics[scale=.44]{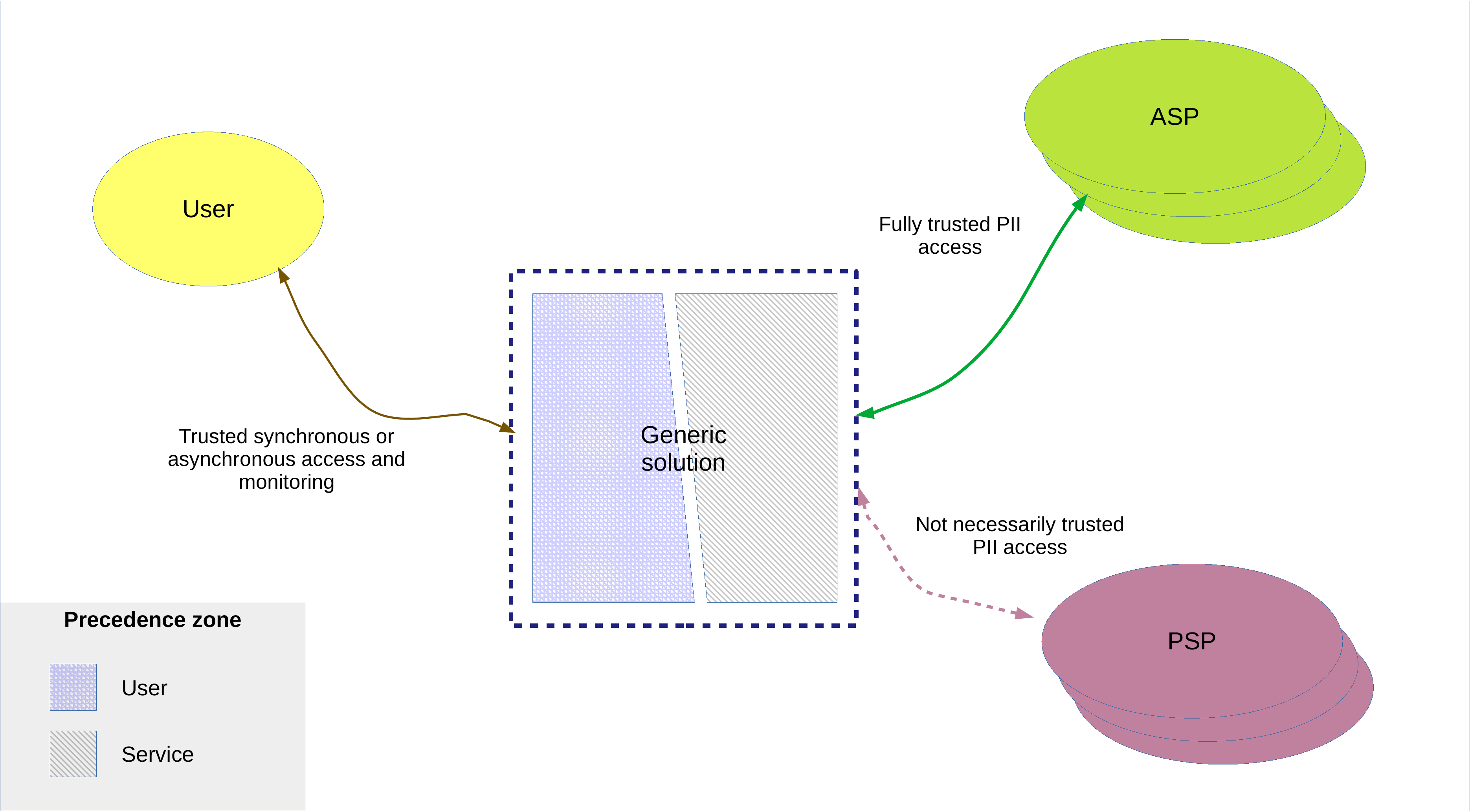}
\caption{Generic architectural layout diagram.\label{generic_arch}}
\end{center}
\end{figure*}

As depicted in Figure~\ref{generic_arch}, the following actors participate in the system:

$\bullet~$ The user is considered trusted and is provided with PII management tools through a UI, in order to help her/him understand the potential risks of her/his PII misuse. \\
$\bullet~$ Administrative Service Provider (ASP) is a trusted entity.
It interacts with the PII self-management solution.\\
$\bullet~$ Private Service Providers (PSP) are semi-honest entities~\cite{Paverd2014}, \textit{i.e.}, they do not try to break the system's technical rules, but instead they try to access any data technically available to them for reading even though they were not functionally meant to be accessed by them.

In this article, we refer to service providers that may be either administrative or private.
Whenever that distinction between administrative and private service providers is not necessary, the acronym SP is used.

\subsection{The territorial collectivity use case in Europe\label{usecase}}
The PII relevant for this use case is documents such as a family register, a copy of an ID card or of a driver license, \textit{etc}.
This information can also be raw PII data such as a string description of a postal address or a geographical location.
Eventually, the metadata linked to these first two types of PII is also considered to be PII.

Our use case takes place in a municipality where the user interacts with the administrative service providers available to her/him for issuing a passport renewal request, as well as for enrolling her/his son in a primary school. \\
Those two requests require one document each, namely a digital copy of the current passport about to expire\footnote{One example is the USA's PassportWizard online service, digitalizing the DS-82 passport renewal paper form, and requiring the user to upload the two pages of her/his current American passport that contain PII.}, and a copy of the family register.

For better convenience, the user also needs to define the set of service providers that will later be able to access these two documents without asking her/him later for proper synchronous (hence blocking) permission (PSPs authorized to access the documents will not wait for the user's permissions, as they already obtained her/his consent).
Supported connected services might be the subscription to local sponsored events or local business offers.
These requirement address the “Tell us once.” French interministerial program.

In the GDPR spirit, the user needs to restrict consent, for both uploaded documents, to a set of purposes.
For better user experience, the user might be proposed a set of purpose categories like administrative procedures, sponsored event subscription, enrollment in local associations, {\it etc}.
Also, once a service provider is revoked, this provider must not illegitimately access the user's PII anymore.

The user also needs to define a validity time window for the authorized access to apply on her/his documents.
He/she can also authorize her/his relatives to use some of her/his documents to ASPs or PSPs.

Eventually, the user also needs to retrieve and manage PII originating from remote PII sources, as this PII may prove necessary while using the administrative and private service providers.

The achieved generic architecture with interactions between users and ASPs/PSPs is depicted in Figure~\ref{generic_arch}.

In summary, the functional requirements for this use case are:\\
$\bullet$ \textbf{Consent management} offered to the user of the SPs (either ASPs or PSPs).\\
$\bullet$ \textbf{Partially-autonomous decision making} to the SPs.\\
$\bullet$ \textbf{Purpose-based authorization} definable by the user.\\
$\bullet$ \textbf{Service provider revocation enforcement} by the user.\\
$\bullet$ \textbf{Time-based authorization validity} defined by the user for any given authorization.\\
$\bullet$ \textbf{Interactions with remote PII sources} through the user-relationship platform.

\subsection{Definitions}
Two core concepts dealing with PII management are introduced below: \\
$\bullet~$ \textbf{Access control} is the enforcement of rules enabling only authorized services to access user data.
Different access control models have been presented in the literature.
These models vary a lot, for instance relying on properties of the requester, or even on the behavior of this requester on the system over time.
\cite{DeCapitanidiVimercati2007} remains accurate in describing the most common AC models.\\
$\bullet~$ \textbf{PII provisioning} is the process of creating or updating data on the user's PII storage base, while PII deprovisioning is the processing of deleting such information.
SPs might be empowered with both provisioning and deprovisioning capabilities.

\section{Selected criteria\label{criteres}}
Tables~\ref{tab:user-gov} and~\ref{tab:data-exchange} give the list of criteria that are of interest for administration and territorial public service providers, among which five, in bold type, are considered as critical for our territorial collectivity use case (see Subsection~\ref{usecase}).
This set of criteria is split into two different categories.

Criteria belonging to the first category deal with user governance, \textit{i.e.}, the ability for users to manage their PII along their whole lifecycle.
Detaining PII (which  denotes a concept wider than the strict physical ownership of PII) gives them the right to control which data processes are applied to their PII.

On the contrary, criteria identifying properties in the data exchange flows that are prone to enforce PII management in a privacy-compliant manner belong to the second category.

\tabcolsep 12pt
\renewcommand\arraystretch{1.3}
\begin{center}
\captionof{table}{User Governance Criteria}\label{tab:user-gov}
\vspace{1mm}
\footnotesize{
\begin{tabular}{l}
\hline\hline\hline
Name of criterion\\
\hline
Type(s) of supported access-control \\
Privacy usability trade-off \\
User interface \\
\textbf{Consent management} \\
Negotiation of data-collection parameters \\
Service provider revocation \\
PII collection purpose definition \\
Inter-user PII sharing \\
Online/offline mode(s) \\
\textbf{Extent of delegation} \\
History/logging of transfers \\
\hline\hline\hline
\end{tabular}
}
\end{center}

\tabcolsep 12pt
\renewcommand\arraystretch{1.3}
\begin{center}
\captionof{table}{Data Exchange Flow Criteria}\label{tab:data-exchange}
\vspace{1mm}
\footnotesize{
\begin{tabular}{l}
\hline\hline\hline
Name of criterion\\
\hline
Type(s) of supported PII \\
\textbf{PII validation} \\
Functional structure \\
Provisioning and deprovisioning management \\
\textbf{Reusability of previously uploaded PII} \\
Minimization management \\
\textbf{Support of remote PII sources} \\
\hline\hline\hline
\end{tabular}
}
\end{center}

\subsection{User governance criteria}
\subsubsection{Type(s) of supported access control}
This AC criterion evaluates how easy, flexible and robust the solution is for the user to elaborate access rules.
These access rules restrict the way her/his PII is shared with ASPs/PSPs.
When correctly defined and applied, they ensure that only legitimate SPs access this PII.
This criterion raises the issues of the underlying security mechanisms and models.
Indeed, these mechanisms and models have an impact on the definable access control policies.
These policies can take the form of an access control list and they can bear more flexible options like contextual information, delegation to an access control agent, cascading authorizations, {\it etc}.

Generally speaking, AC encompasses both user and SP access control.
However, as users are considered trusted in our system model and requirements -- see Section~\ref{fundamentals} --, efforts shall focus on the SP access control, especially in our model since the consequences of SPs illegitimately accessing PII may be significantly harmful.

\textbf{Reason for criterion selection.}
This criterion directly addresses the extent of possibilities for the user to define access control over her/his PII, thus respecting the territorial use case presented in Subsection~\ref{usecase}.
The access control capabilities indeed have a significance regarding the extent of the ability for the user to manager her/his own PII when dealing with her/his online territorial services.

\subsubsection{Privacy usability trade-off}
This criterion defines whether privacy enforcement happens at the expense of the usability of the solution.
Usability and privacy are sometimes viewed as contradictory objectives in identity management.
However, depending on the nature of the solution, such a trade-off can be softened or sometimes even avoided.
For instance, some locally-deployed user-centric solutions may enforce privacy without degrading the usability of these solutions (due to proximity of the user and the deployment on trusted hardware).

User privacy is enforced towards the SPs.
For some selected solutions, user privacy also needs to be enforced towards the solution -- when, under some security hypothesis, it cannot be trusted.

This criterion raises several concerns.
First, the users should understand the privacy-enforcement mechanisms and the reason for the trade-off.
Second, they should know their privacy-related rights while using the services -- and the solution should guide them to the enforcement of these rights.
Third, the enforcement of privacy-related rights should not happen at the expense of the functional coverage of the solutions.
Fourth, an adjustable privacy-usability trade-off might be of interest, but then remains the question of identifying which actors of the use case would be able to configure or adjust that trade-off.

\textbf{Reason for criterion selection.}
The solutions enforcing the territorial use case are managed by the user.
Respecting privacy properties must be essential to them, but the enforcement of these properties should not come at the price of usability reduction.
Indeed, a significant trade-off reduces the solution's efficiency, thus defeating the use case.

\subsubsection{User interface}
This criterion measures how efficient and user-friendly the user interface is.
It involves studying the understandability of the supported functionalities by the users, that is how it helps the user understand how her/his privacy and security are ensured.

\textbf{Reason for criterion selection.}
The fact that the selected solutions are user-driven of course implies the presence of a user interface.
This user interface directly conditions the possibilities of empowerment of the user when dealing with the solutions.
Studying the UI for each solution is therefore necessary.

\subsubsection{Consent management\label{crit-consent-management}}
Evaluating solutions according to this criterion implies the study of the user consent for these PII exchanges.
Thus, in this particular context of potentially sensitive PII collection for later processing, the consent should be perceived as an agreement granted to a service by the user.

\textbf{Reason for criterion selection and criticity.}
This criterion is tagged as critical as it is a fundamental part of user's privacy guarantees: it supports the margin of decision that the user has upon the transfer of her/his PII from or to SPs.
Indeed, according to our territorial use case in Subsection~\ref{usecase}, the user's PII may be used by ASPs as well as PSPs, for various purposes, and therefore require proper consent management.

\textbf{Underlying technical challenge.}
The technical challenge for this criterion is the ability for the solution to handle a consistent consent model, suitable for the administrative use case.
In other words, the consent model should be suitable for managing PII all along its lifecycle and tackle the subset of services (\textit{e.g.}, social and family-related services, health-related services and civil procedures) that the public administration provides.
From a technical point of view, it means that the consent model has to bear (i) spatial information, \textit{i.e.}, the extent to which the consent is given and the services allowed to access the user's PII, as well (ii) temporal information, \textit{i.e.}, the time at which the consent was given, its the expiration date, and possibly version numbers of the end-user licence agreement for the services being granted authorizations.

\subsubsection{Negotiation of data collection parameters}
We consider the agreement -- between a user, owning her/his PII, and a service provider requesting PII -- to be a contract.
The parties may be able to negotiate the terms of the agreement before the agreement is met.
If so, the user and the service should be able to define, in an interactive way, the conditions of PII collection.
This means that the solutions should enable the user to define which PII a service may be able to collect.
In return, the solution should let the service declare guarantees of a correct PII usage:
the user should obtain effective data-processing parameters from the service (such as the data-retention period, the extracted PII or the final use of PII) in exchange of the consent to the collection of her/his PII.
Other collection parameters may be supported, such as the precision of the PII sent, depending on the recipient's legitimacy -- \textit{i.e.}, the user being able to accept a less legitimate SP's request, provided that the PII sent go through some noise-introduction algorithm beforehand.

Supporting such a negotiation is a way to meet the requirements of the GDPR, such as the principles of finality\footnote{Any PII collection should happen for a specific purpose.} and the minimization\footnote{The SP should not collect more PII than what is necessary to achieve its declared processing.} of PII collection.

Additionally, we should question whether the negotiation can be initiated on a mutual trust basis between the two parties, and more particularly we should wonder whether the solution: \\
$\bullet$ is able to define a (quantitative or qualitative) level of trust granted to the service. \\
$\bullet$ lets the user adjust that trust level. \\
$\bullet$ supports synchronous negotiations.\\
$\bullet$ supports asynchronous negotiations.

\textbf{Reason for criterion selection.}
The use case states that the user should be able to define which PII is collected by the ASPs and the PSPs interacting with the user-relationship management platform.
However, the PII initially claimed by these SPs may not be a subset the set of PII that the user initially agrees to disclose.
The only solution to this common case is the enforcement of a negotiation process.

\subsubsection{Service provider revocation}
The enforcement of PII access revocation for a SP should be implemented.
For instance, a user having previously given consent to a PSP for accessing the copy of her/his ID card should be able to revoke the access at any time.
Regarding this example, the access grant was for instance given to the SP to validate users subscription to a local sports event.
Once the sports event is over, the user should be able to revoke the PSP authorized access to the copy of each user's ID card.
The PII management system should provide the user with a simple way to perform that operation.

\textbf{Reason for criterion selection.}
On top of being an obviously-expected feature, the ability to revoke service providers is a legal requirement: according to the GDPR, and more especially its right to oppose and its right for correction, the user must be able to change previously-given consents at any time.

\subsubsection{PII collection purpose definition}
This criterion deals with the ability for the user to set, or at least to validate, the authorized purposes for which her/his PII can be collected.
For instance, the user may want that a copy of her/his ID card be limited to administrative purposes only, or, on the contrary, extended to private services.
The solution should address that need.

\textbf{Reason for criterion selection.}
Once again, the ability for the user to restrict the PII collection to a specific and explicit set of purposes is a legal requirement.
Purpose-based authorization is also a convenient access-control abstraction, easier for the user than per-case SP authorization.

\subsubsection{Inter-user PII sharing}
Sharing PII implies that a user Alice is able to share a document, \textit{e.g.}, her family register, with another user Bob, for legitimate reasons -- for instance because, as a family member, his identity also appears on that document.
Hence, when using services that require Bob to supply his family register, he may be able to use the shared document instead of supplying another copy.
We select this criterion to determine whether the solutions address this need.

\textbf{Reason for criterion selection.}
Despite not being a critical criteria, the ability to share PII between user can be considered as an elegant way of enforcing some parts of the use case -- especially when some of the involved PII describes several users at the same time.

\subsubsection{Online/offline modes\label{on-off}}
This criterion addresses the user's control over her/his PII exchanges, whether she/he is ``present'' when these exchanges happen.
A supported offline mode means that the user can configure a software agent to manage PII exchanges happening afterwards, while she/he is offline.

An online-mode-only solution requires the user to synchronously manage the exchange and collection of her/his PII.
This means that the system shall wait for the user's next connection, in order for her/him to validate the PII exchange.
As a consequence, an online mode, when implemented alone, makes it difficult to ``industrialize'' the solution, as automatic PII exchanges between services are not supported.

\textbf{Reason for criterion selection and criticity.}
This criterion is tagged as critical as the territorial use case emphasizes the ability of the solution to take actions on behalf of the users when they are online.

\textbf{Underlying technical challenge.}
The underlying technical challenges of the support of online and offline modes are as follows:
\begin{itemize}
\item Providing the solution with a fine-grained consent model that is able to support both modes.
\item Enforcing traceability of operations, especially while in offline mode.
\item Providing users with an interface which exposes the capabilities of these two modes.
\end{itemize}

\subsubsection{Extent of delegation\label{crit-extent-delegation}}
The autonomous behavior of the solutions consists in authorizing, on behalf of the user (whether she/he is online or not), some SPs to perform operations on elements of the user's PII data base.

Therefore, the user may want to define the degree to which the PII management solution should act on her/his behalf.
The user should be able to require the solution to ask for her/his explicit consent each time an SP wants to access some of her/his PII.
Conversely, the user could decide that the solution acts autonomously on her/his behalf once a delegation policy has been set.\\

The previous criterion, presented in Subsection~\ref{on-off}, deals with temporal autonomy of the selected solutions, \textit{i.e.}, allowing the study of asynchronicity properties of the selected solutions.
On the contrary, this criterion deals about these solutions' spatial autonomy, \textit{i.e.}, the ability for the selected solutions to act on behalf of the users at different steps of the management of their PII.

\textbf{Reason for criterion selection and criticity.}
Ensuring this delegation means finding the right balance between assistance and consent, \textit{i.e.}, the user should be assisted by the delegation process to offload her/his decisional tasks, while her/his consent should be respected.
Eventually, this criterion is tagged as critical for our use case, as a consistent delegation model is necessary to take care of PII transfers on behalf of the user.
As depicted in our use case in Subsection~\ref{usecase}, the purposes for ASP or PSP may vary a lot.
Instead of defining fine-grained management of their PII, users may prefer to define delegation policies for the system to next adapt their behavior when dealing with ASPs and PSPs.

\textbf{Underlying technical challenge.}
The technical challenge of these modes is providing the user with the ability to define which PII processing should happen when she/he is offline.
The consent model proposed by the solution should be able to apply the user's choices regarding offline PII processing.
From a technical point of view, the spatial/temporal distinction defined in Subsection~\ref{crit-consent-management} still applies for the extent of delegation.
The delegation operates spatially.
As a result, the following questions must be answered when studying the solution according to this criterion:
\begin{itemize}
\item Does the solution provide a categorization of services, allowing generic consent to be given by the user?
\item Similarly, is PII categorization supported by the solution?
\item Does the solution define fine-grained types or scopes of actions for the consent given by the user?
\end{itemize}

\subsubsection{History/logging of transfers}
This criterion enforces the ``right to be informed'' and enables the users to access the history of all executed transfers, with several possible granularities, \textit{e.g.}, from the PII metadata only, to the full data content.

It raises the following concerns:\\
$\bullet$ the need for the user to understand which PII have been exchanged, and when it had happened. \\
$\bullet$ the quantity of information revealed by the log or history facility about the content of the PII exchanged.

\textbf{Reason for criterion selection.}
The logging of transfers is also considered as a legal requirement, enforcing the principle of privacy by design stated by the GDPR.

\subsection{Data exchange flow criteria}
\subsubsection{Type(s) of supported PII\label{def-pii-types}}
The three following types of PII are considered:
\begin{itemize}
\item User structured {\bf documents}, \textit{e.g.}, a digital copy of a family register.
\item Raw {\bf data} (usually under the form of non-document data formats, such as XML or JSON, or plaintext), \textit{e.g.}, the user's postal address or date of birth.
\item {\bf Metadata}, \textit{i.e.}, attribute or value metadata, for instance as specified in NIST internal report 8112~\cite{NISTIR8112} for federated identity systems.
\item[]\end{itemize}

These types of PII lead to support several possible use cases.
For instance, the solution may be used as a platform for storing personal and secure storage of documents owned by the user.
The deployment of such a tool may be performed on a remote Web server, or on a local dedicated hardware (on a phone, tablet, SoC, {\it etc.}).

With metadata, a higher abstraction level is provided, thus leading to a number of possible use cases including validation based on data content, time-to-live metrics, data type, identity, {\it etc}.

\textbf{Reason for criterion selection.}
The supported PII types are properties of interest as they are often determined by the SP for its own usage.
They also give a hint of how much governance is left to the user regarding her/his own data.

\subsubsection{PII validation\label{crit-pii-validation}}
Validated data might be required by some services.
For instance, a sport association in the user's municipality may require a certified digital copy of the user's ID card, as a necessary document for the enrollment process.

The objective is to mitigate identity theft by avoiding the use of false or stolen PII.
This criterion expresses whether such a validation process is supported by each of the selected solutions.
We purposely do not give a precise definition of this validation, as it can encompass several processes, such as the manual validation by a human operator of the territorial collectivities or the public administration, or even a validation performed by a trusted authority.

\textbf{Reason for criterion selection and criticity.}
This criterion is tagged as critical as it is necessary for collectivities and administrations in order to perform an efficient overall validation of procedures.
Indeed, PII validation highly enables partial automation of the procedures, which spares the human agent from repetitive tasks -- the latter can in return provide help regarding more complex or non automatable work.

\textbf{Underlying technical challenge.}
From an administrative service's point of view, validating PII at different stages of their processing enables different workflows.
To some extent, validated PII makes it possible to skip some processing states in the administrative functional workflow.
For instance, validated PII may not need a manual verification from administrative agents whereas invalidated PII may.
The technical capabilities of the solution must be able to follow the functional scenarios of PII validation.
From a technical point of view, this criterion raise the following questions:
\begin{itemize}
\item Does the origin of the PII (for instance when this PII may come from any of several PII sources) takes part in determining the degree of validation of this PII?
\item Can the solution include different validation scenarios depending on the lifecycle stage at which the PII is?
\end{itemize}

\subsubsection{Functional structure}

The functional structure determines the roles of components within the architecture, identifying the requirements for the user to be able to manage her/his PII with the solutions according to the use case.
This criterion is qualitative.
The structural variations between the solutions is too wide, hence no quantitative metrics is used for this criterion.

However, from a qualitative point of view, the different functional components of each solutions is a useful indication of what functionalities are offered to the user.
The components' layout reveals the degree to which the solution can be managed by the user.
For instance, some solutions may provide a user-managed agent, acting as an access point to the management of the solution by the user.

On the contrary, some other solutions implement functional components -- such as third-party entities, arbiters, {\it etc.} -- that do not necessarily abide by the user's wishes in terms of PII management.

\textbf{Reason for criterion selection.}
As explained in the previous paragraphs, the functional structure gives information about the supported elements of the territorial use case.
However, it cannot be considered as critical.

\subsubsection{Provisioning and deprovisioning management\label{crit-provdeprov}}
This criterion deals with the possibility for PSPs or ASPs to act as a PII source for the user.
Such a criterion is particularly relevant in case of administrative or territorial service providers \footnote{The French Unique reglementary act RU030 defines a set of public online services for which the collection and provisioning of some PII is legitimate.
This text of law acknowledges the importance of such services of public interest, while ensuring that no unnecessary data is collected or provisioned.
See \url{https://www.legifrance.gouv.fr/loda/id/JORFTEXT000027697207/2021-01-19/} (\textit{French resource}; last accessed: March 10, 2021).}.
Such administration (or territorial collectivity) detains sensitive PII describing their users, and their provisioning needs to be managed.
This may also be necessary for PSPs.

In order to define and qualify this criterion, we express the following concerns: \\
$\bullet$ the ability for the ASP or PSP to provision data on behalf of the user. \\
$\bullet$ if so, the provisioning taking place on the storage entity directly receiving PII from the user. \\
$\bullet$ optionally, the precedence of authority (between the user and the SP), regarding the provisioned PII.

This criterion also tackles the management of PII deprovisioning on the solution, \textit{i.e.}, the planned deletion of PII, when reaching the end of its lifecycle.
Managed deprovisioning involves respecting a main concern of data protection, the ``right to be forgotten'', defined by the recent European regulation as the right to erase its own PII.

It means that the user should be able to make some PII be unusable to SPs anymore --  the most obvious way to do so being PII deprovisioning on these SPs.

\textbf{Reason for criterion selection.}
Provisioning and deprovisioning capabilities are a direct part of the use case as they enable SPs to perform operations on the user's PII after obtaining her/his consent to do so.

\subsubsection{Reusability of previously uploaded PII\label{reusability}}
This criterion means that the user can reuse some PII previously uploaded or provisioned, in order to fulfill another later processing.
For instance, after uploading a copy of her/his family register so as to enroll her/his son in primary school, the user may have to use this document again in order to enroll her/his daughter in the municipality's day nursery (see Subsection~\ref{usecase}).
This also avoids that the user loads the same PII twice into the system.

\textbf{Reason for criterion selection and criticity.}
The “Tell us once program” is a core element of the use case (see Subsection~\ref{usecase}), as it is directly relevant to the reusability dimension.
This is a critical criterion.

\textbf{Underlying technical challenge.}
The technical challenge of PII reusability involves determining when the conditions for such reusability are gathered.
These conditions are:
\begin{enumerate}
\item Enforcing user consent.
In other terms, this challenge implies studying whether the evaluated solutions support PII reusability without undermining user consent.
\item Applying the legal framework of PII processing, as stated by the GDPR.
In other terms, we need to know whether the reusability is adaptable to the legal framework applied in the territorial collectivity.
\item Applying applicative rules specific to the context of administrative services.
In other words, the evaluation according to this criterion need to identify the solutions supporting existing variations in such reusability rules.
\end{enumerate}

\subsubsection{Minimization management}
Respecting this criterion guarantees that the ASP and PSP are collecting PII in a minimized way, \textit{i.e.}, only the PII strictly needed for the data processing as declared by the SP.
For instance, a PSP only requiring the user's name and postal address should not access the entire content of her/his ID document (such as the document serial number, the place of birth, \textit{etc.}).

\textbf{Reason for criterion selection.}
The principle of finality of the European GDPR results in the minimization of the PII collected by the ASPs and the PSPs.
It is a legal requirement, but is not considered functionally critical considering the territorial use case.

\subsubsection{Support of remote PII sources\label{crit-remote-sources}}
Users managing their PII may want the solution to abstract the differences appearing while managing remote PII sources.
As a result, supporting this criterion ensures that the solution provides such an abstraction regarding the PII location.
Whether the PII is locally stored on the solution or remotely accessible on remote sources shouldn't hinder the user in her/his PII management.\\

This criterion is tagged as critical as some procedures in our use case require PII that are available on remote sources only.
The use case in Subsection~\ref{usecase} involves the use of remote PII sources, for instance the ones supplied by the French administration, such as FranceConnect identity providers~\footnote{See \url{https://franceconnect.gouv.fr/} (last accessed: March 10, 2021).}, the DGFiP (central fiscal system) data source and the CNAF (children and family allowances) data source.

\textbf{Reason for criterion selection and criticity.}
The support of remote sources is considered critical as it is at the core of the use case relying on several sources maintained by official authorities at national and European levels.

\textbf{Underlying technical challenge.}
The underlying technical challenge involved in the support of remote PII sources is threefold:
\begin{enumerate}
\item Consent management must apply in a consistent manner regardless of PII origin.
That is, the consent model enforced must be compatible with all the sources.
\item Sources trust management must be enforced, as somes sources may be more reliable or more honest than others.
For instance, some sources may act in a honest-but-curious manner, gathering information that wasn't meant for them to collect.
\item Last but not least, interoperability concerns arise when dealing with several sources, with potential variations in access and authorization protocols.
\end{enumerate}

\section{Taxonomy and list of technical approaches\label{taxonomy}}

Thirteen solutions are classified into two families and four different categories, as depicted in Figure~\ref{arbre}.

{\tiny
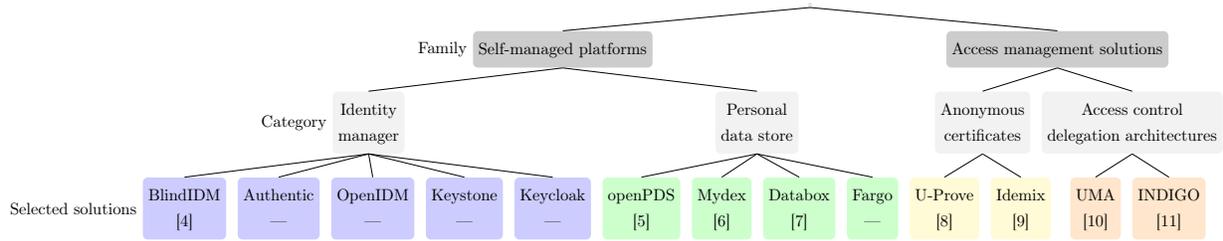
\begin{figure*}
\centering
 \scalebox{.62}{
\begin{forest}
  for tree={
    align=center,
    parent anchor=south,
    child anchor=north,
    l sep=5mm,
    my blue box,
  }
  [, circle, fill, inner sep=1pt
    [Self-managed platforms, label=left:Family, fill=black!20
       [Identity \\
      manager, label=left:Category, fill=black!5
        [BlindIDM\\ \cite{Nunez2014}, label=left:Selected solutions, fill=blue!20]
        [Authentic\\ ---, fill=blue!20]
        [OpenIDM\\ ---, fill=blue!20]
        [Keystone\\ ---, fill=blue!20]
        [Keycloak\\ ---, fill=blue!20]
      ]
     [Personal \\
      data store, fill=black!5
        [openPDS\\ \cite{10.1371/journal.pone.0098790}, fill=green!20]
        [Mydex\\ \cite{Papadopoulou2015}, fill=green!20]
        [Databox\\ \cite{Mortier:2016:PDM:3010079.3010082}, fill=green!20]
        [Fargo\\ ---, fill=green!20]
      ]
    ]
    [Access management solutions, fill=black!20
      [Anonymous \\
      certificates, fill=black!5
        [U-Prove\\ \cite{u-prove-technology-overview-v1-1-revision-2}, fill=yellow!20]
        [Idemix\\ \cite{Camenisch2007}, fill=yellow!20]
      ]
      [Access control \\
      delegation architectures, fill=black!5
        [UMA\\ \cite{UMASpecs}, fill=orange!20]
        [INDIGO\\ \cite{CeccantiINDIGO}, fill=orange!20]
      ]
    ]
  ]
\end{forest}
}
\caption{Taxonomy of PII management solutions with their categories, partitioned in two main families.}
\label{arbre}
\end{figure*}
}

The first family of solutions (\textit{i.e.}, self-managed platforms) encompasses monolithic software entities that provide users with PII management capabilities. \\
On the contrary, the second family of solutions (\textit{i.e.}, access management solutions) is set at a higher architectural level, and can't be identified as a single software tool.
The interactions among several entities of solutions of that latter family enable PII management capabilities for users.

These two families then form a complete partition of PII management solutions.
Each of these families is split in two categories for further disambiguation.
The four resulting categories are presented in the following paragraphs.

All these solutions have been selected for their detailed presentation in published scientific articles, or for their thorough documentation and code available under a free license.
This figure describes both academic (BlindIdM, openPDS, Databox, Idemix, U-Prove and INDIGO) and industrial (Mydex, Fargo, Authentic, OpenIDM, Keystone, Keycloak, and UMA) solutions.
Industrial solutions are considered as targeting operational ground, and thus can be expected to cover a wider set of properties of interest than academic ones.

As shown in the zoom-in diagrams of Figure\ref{generic_arch}, \textit{i.e.}, Figures~\ref{idm_arch} --- \ref{deleg_arch} for the four categories of solutions, these solutions -- although covering similar functional objectives -- have different architectural and functional natures, with different acting entities.
These observations serve to establish our arborescent taxonomy, with the two following identified families.

Family - self managed platforms - contains two categories of solutions:
\begin{itemize}
\item Personal Data Stores (PDS), for solutions providing a service for PII storage to the user.
\item Identity Managers (IdM), which often double as identity providers in the context of federated identity management (FIM) and single sign-on (SSO).
They help the user handle her/his personal accounts and her/his associated identity information.
\end{itemize}

Family - access control management - also contains two different categories:
\begin{itemize}
\item Anonymous certificates, meant for the user to enforce data minimization, \textit{i.e.}, reveal to the SP only the PII strictly necessary for data processing.
\item Access control delegation platforms, meant for the user to define fine-grained authorization on her/his PII.
\end{itemize}

\subsection{Identity Managers (IdM):\label{idm-pres} {\it (BlindIDM, Authentic, OpenIDM, Keystone, Keycloak)}}

Identity managers (IdMs) offer an endpoint for the user-driven management of her/his digital identities, especially when provided across SPs (see Figure~\ref{idm_arch}).

\includegraphics[width=\columnwidth]{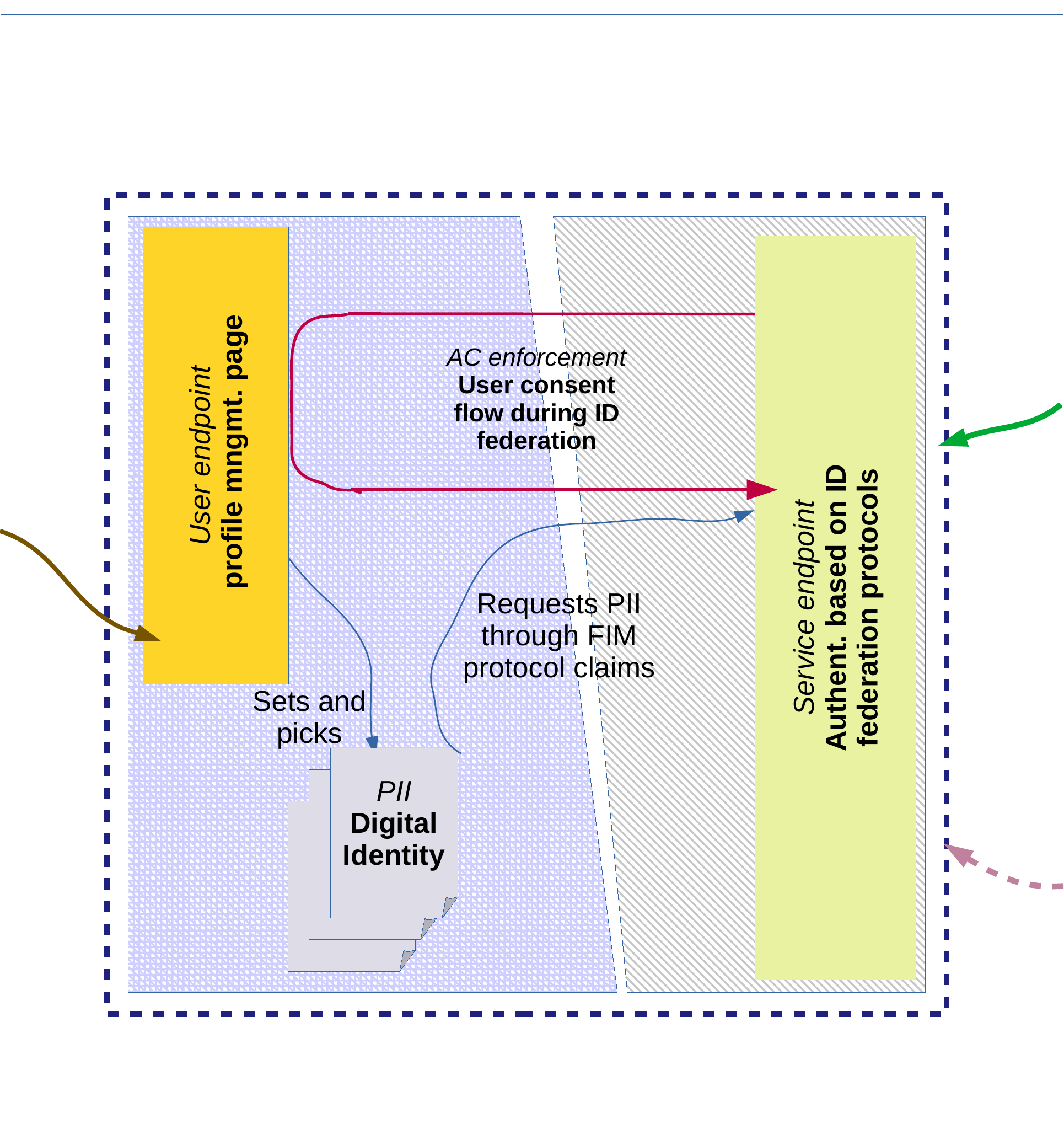}
\captionof{figure}{IdM architectural layout diagram.\label{idm_arch}}

Depending on the implementation and the configuration, some PII management capabilities are offered to the user while others are left to the administrator only.
For example, the user may be able to manage the services requesting her/his PII, while the administrator may only declare the exact set of services connected to the IdM at any time.

In addition, the selected IdMs also bear the role of identity providers within a federated identity management~\cite{Camenisch2007} (FIM) system, helping the user handle a set of common identities among several registered services.
FIM is performed using standardized protocols -- either SAML~\cite{RFC7522} or the OpenID Connect identification protocol~\cite{OIDCCore} derived from the OAuth2 authorization framework~\cite{RFC6749}.
This framework provides a clear decorrelation of roles when providing access management.
These roles are respectively the client, the relying party, the authorization server and the resource server.
A system of grant types enables different authorization flows each supporting some specific client types and security hypotheses (\textit{e.g.}, the ability to share secrets, the ability for the client to store its own client secret, \textit{etc.}).
Although originally designed as a lightweight protocol -- in contrast, for instance, with SAML --, OAuth now specifies richer functionalities such as the dynamic registration of clients~\cite{RFC7591}, the dynamic management of clients~\cite{RFC7592}, protocol interoperability through its assertion framework~\cite{RFC7521}, server metadata~\cite{RFC8414} and token introspection~\cite{RFC7662} and revocation~\cite{RFC7009} endpoints.
OIDC also specifies backchannel authentication flows~\cite{OIDCBackchannelAuthn}, to ``compete'' with SAML (SOAP-based) artifact-resolution bindings.

The FIM property of interest for this survey is the Single Sign-On (SSO), \textit{i.e.}, the ability for an identity provider to maintain a single authenticated session across several SPs, and its complementary property is Single Logout (SLO).

A basic layout of FIM SSO authentication flow is shown in Figure~\ref{fig:fimflow}, where the client is redirected by the federated service to the identity provider for authentication and for the obtention of an authorization token.
In case of IdMs using exclusively Web technologies, the specified data formats are XML-based or JSON-based, communications are TLS encrypted and the client application is a simple browser.

\vspace{5mm}
\begin{center}
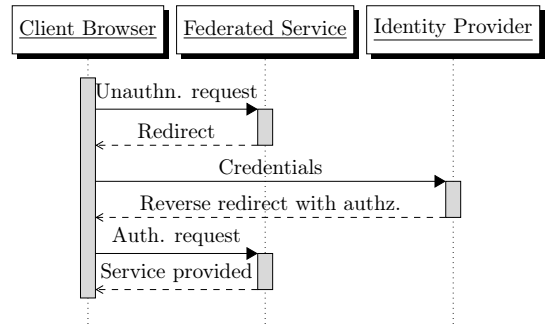

\scalebox{0.8}{
\begin{sequencediagram}

    \newthread{A}{Client Browser}{}
    \newinst{B}{Federated Service}{}
    \newinst{C}{Identity Provider}{}

    \begin{call}{A}{Unauthn. request}{B}{Redirect}{}
    \end{call}
    \begin{call}{A}{Credentials}{C}{Reverse redirect with authz.}{}
    \end{call}
    \begin{call}{A}{Auth. request}{B}{Service provided}{}
    \end{call}

\end{sequencediagram}
}
\captionof{figure}{FIM authentication sequence diagram.}\label{fig:fimflow}
\end{center}

IdMs support a common core of functions, but they have their own management particularities.
The basic common core of functions are authentication and user profile management.
The IdMs hence support the usual authentication challenges over HTTP(S) (Basic, Digest) as well as token-based authentication schemes (Bearer).
User profile management involves the ability for the user to set PII profile attributes to be federated over the different SPs registered in the federation.
This common core of functions requires that the IdMs support certain set of protocols involving public-key cryptography~\cite{Diffie1976}, secret sharing~\cite{Shamir1979} and cryptographic one-way hash functions~\cite{Schneier_appliedcryptography}.
These technical requirements will however not be discussed further in this article.

\textbf{\textit{Selected solutions}}

This category includes the following solutions: BlindIDM~\cite{Nunez2014},
Authentic\footnote{See \url{https://dev.entrouvert.org/projects/authentic} (last accessed: March 10, 2021).},
OpenIDM\footnote{See \url{https://backstage.forgerock.com/docs/openidm} (last accessed: March 10, 2021).},
Keystone\footnote{See \url{https://docs.openstack.org/keystone/pike/} (last accessed: March 10, 2021).} and
Keycloak\footnote{See \url{https://www.keycloak.org/} (last accessed: March 10, 2021).}.

BlindIDM is an IdM whose deployment is meant for untrusted (semi-honest) servers.
As a reencryption proxy -- as for instance presented in~\cite{Blaze98atomicproxy} --, it handles PII without reading its content, thanks to a set of asymmetric reencryption keys.
These reencryption keys are used to translate a ciphertext encrypted with the user's private key, into a ciphertext encrypted with the SP's private key, with no knowledge of the original cleartext message.

Authentic is an identity provider focused on modularity and extensibility to a wide number of identity management protocols.
This modularity is achieved by using the Django Web framework.

OpenIDM, Keycloak and Keystone are three Web identity management servers, presenting similar functional coverage.
Keycloak and Keystone are the main identity management tool maintained by RedHat, and the identity management layer of the OpenStack cloud-computing framework project respectively.
While OpenIDM and Keycloak are thorough generic-purpose identity managers, Keystone is used as an identity-management interface provider among the different services of Openstack.
These three solutions are solving a certain number of well-known concerns of identity management, such as access control, PII provisioning, data reconciliation, password management, and so on.
They are designed for a deployment in wide scalable software ecosystems, possibly in federated identity environments.

\subsection{Personal Data Stores (PDS): {\it (openPDS, Mydex, Databox, Fargo)\label{pds-pres}}}

PDSs store data on behalf of the user either locally -- on user hardware -- or not.
PDSs are split into a typical set of functional components (see Figure~\ref{pds_arch}):
The data store is separated from the user front endpoint.
The user can access her/his PII data and documents through the user interface.
SPs requests are filtered by an Access Control module which is responsible for granting access, in a partially automated manner (at least), to legitimate providers only.
The Access Control module is managed by the user who is typically maintaining a list of legitimate SPs.
The user also decides when and how PII can be collected and processed.

\textbf{\textit{Selected solutions}}

The different solutions for this category are openPDS~\cite{10.1371/journal.pone.0098790},
Mydex~\cite{Papadopoulou2015},
Databox~\cite{Mortier:2016:PDM:3010079.3010082} and
Fargo\footnote{See \url{https://dev.entrouvert.org/projects/fargo} (last accessed: March 10, 2021).}.

openPDS enables the user to adapt the accuracy of the answers provided to the SP.
This solution, along with its SafeAnswers framework, is designed to receive questions (\textit{i.e.}, algorithms) from the SPs meant to be run locally against the user's PII metadata.
Thus no users' raw PII metadata is sent to the SPs.
As a result, only the output of the algorithms (considered as the ``safe'' answers to the questions) is known by the SPs.

Mydex offers classic personal data storage features to the user.
It contains a simple access control interface which enables the user to configure her/his temporal and spatial access control -- ``temporal'' as a time window is defined for the validity of the PII access grant, and ``spatial'' as a list of registered services is authorized to access the data.

Databox proposes a privacy-preserving interoperable multi-component PDS architecture.
The architecture is centralized and defines an arbiter responsible for the communications between the components.
It is meant to be deployed locally (on user hardware, for instance), with possible access from remote servers.

Fargo acts as a simple document storage server.
As a Web application, its modular structure makes it connectable to other components, such as identity providers and data consuming service providers.
It is meant for limited straightforward use cases (documents reusability in user forms) and does not implement complex features.

\includegraphics[width=\columnwidth]{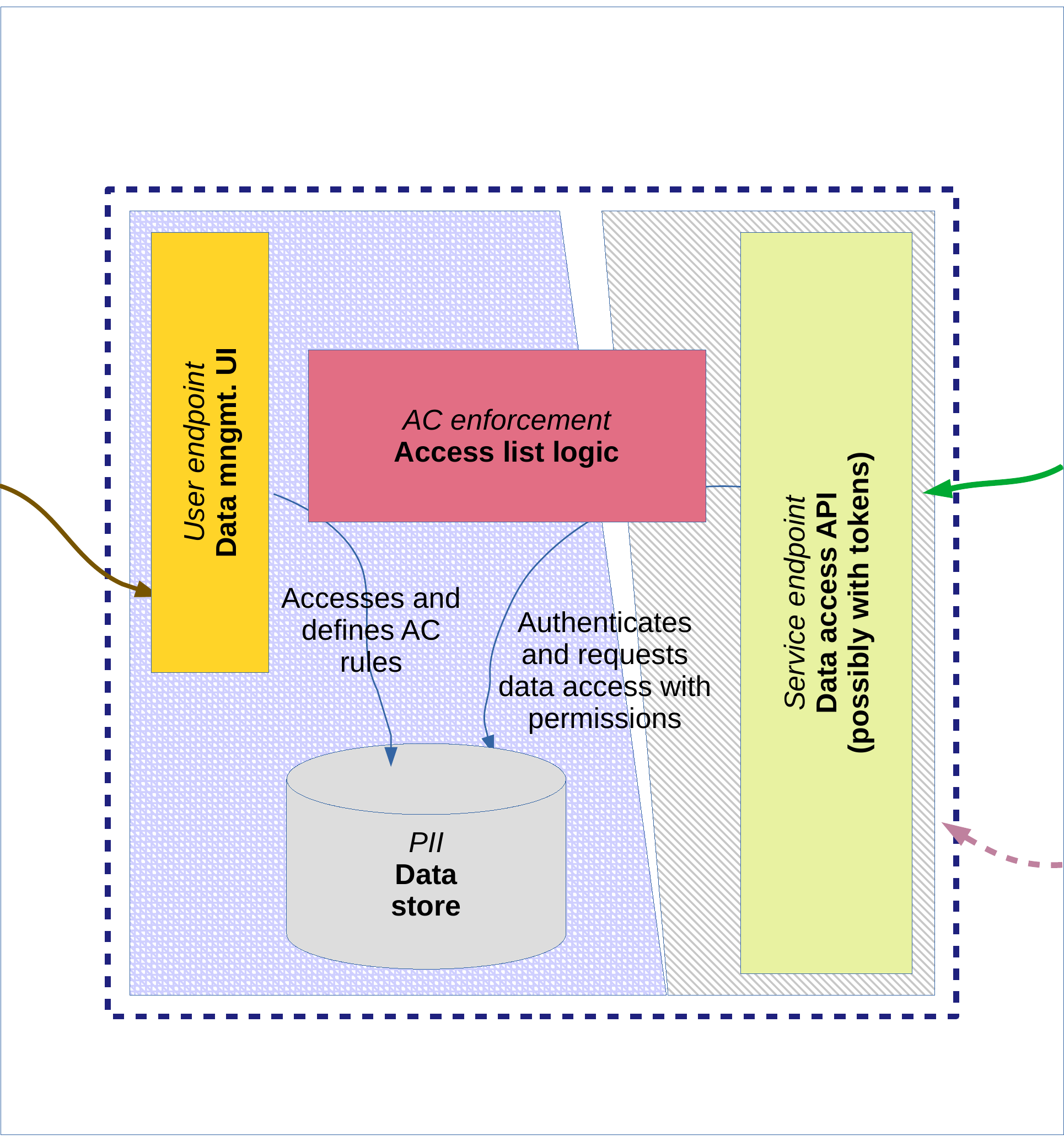}
\captionof{figure}{PDS architectural layout diagram.\label{pds_arch}}

\subsection{Anonymous certificate systems:\label{anoncred} {\it (U-Prove, Idemix)}}

These solutions rely on a four-party architecture (see Figure~\ref{anoncred_arch}) to support the certification of user attributes (as well as properties derived from these attributes).
In these solutions, a Prover (P) owns some PII.
(P) obtains anonymous certificates from an Issuer (I).
(P) is able to prove the validity of some (properties over) PII to a Verifier (V) which is the PSP in our scenario.
Depending on some contractual conditions, a Revocation Referee (RR) may cancel the anonymity and thus re-identify the certificates -- \textit{i.e.}, identify the owner of the certificates.

The anonymity property is ensured thanks to the two following principles: the communication unlinkability and the zero-knowledge proofs of knowledge~\cite{Blum1988,Chaum1987}.

Communication unlinkability refers to the security protocol preventing any entity over the network from possibly determining whether two proofs have been issued for the same user -- even the destination services, acting as verifiers.
Moreover, the proofs, which can be established to validate some PII or some properties over this PII, are benefiting from the zero-knowledge property: no additional information -- apart from the veracity of the proof -- can be inferred from the proof verification process.

These solutions are relevant for self-consistent transactions, in which the SP does not need to be bound to a specific user or to any other transactions.
For instance, in case of online games, the PSP has to check whether a user is an adult before granting her/his access, but it does not have to learn about her/his name, her/his date of birth, {\it etc}.
As such, only the information strictly needed is collected by PSP.
Additional properties can also be ensured, such as non-replay of proofs, which is of interest for specific scenarios (\textit{e.g.}, electronic cash transactions).

\textbf{\textit{Selected solutions}}

The two selected solutions for this category are Idemix (``identity mixer'')~\cite{Camenisch2007} and U-Prove~\cite{u-prove-technology-overview-v1-1-revision-2}.
They both propose similar features, \textit{i.e.}, a set of anonymous certificates for users to freely use with SPs.
These certificates enforce: \\
$\bullet$ service unlinkability: a service $A$ is not able to determine that the user is also served by a service $B$. \\
$\bullet$ transaction unlinkability: a service $A$ is not able to determine whether two of its transactions have been issued for the same actual user.

\includegraphics[width=\columnwidth]{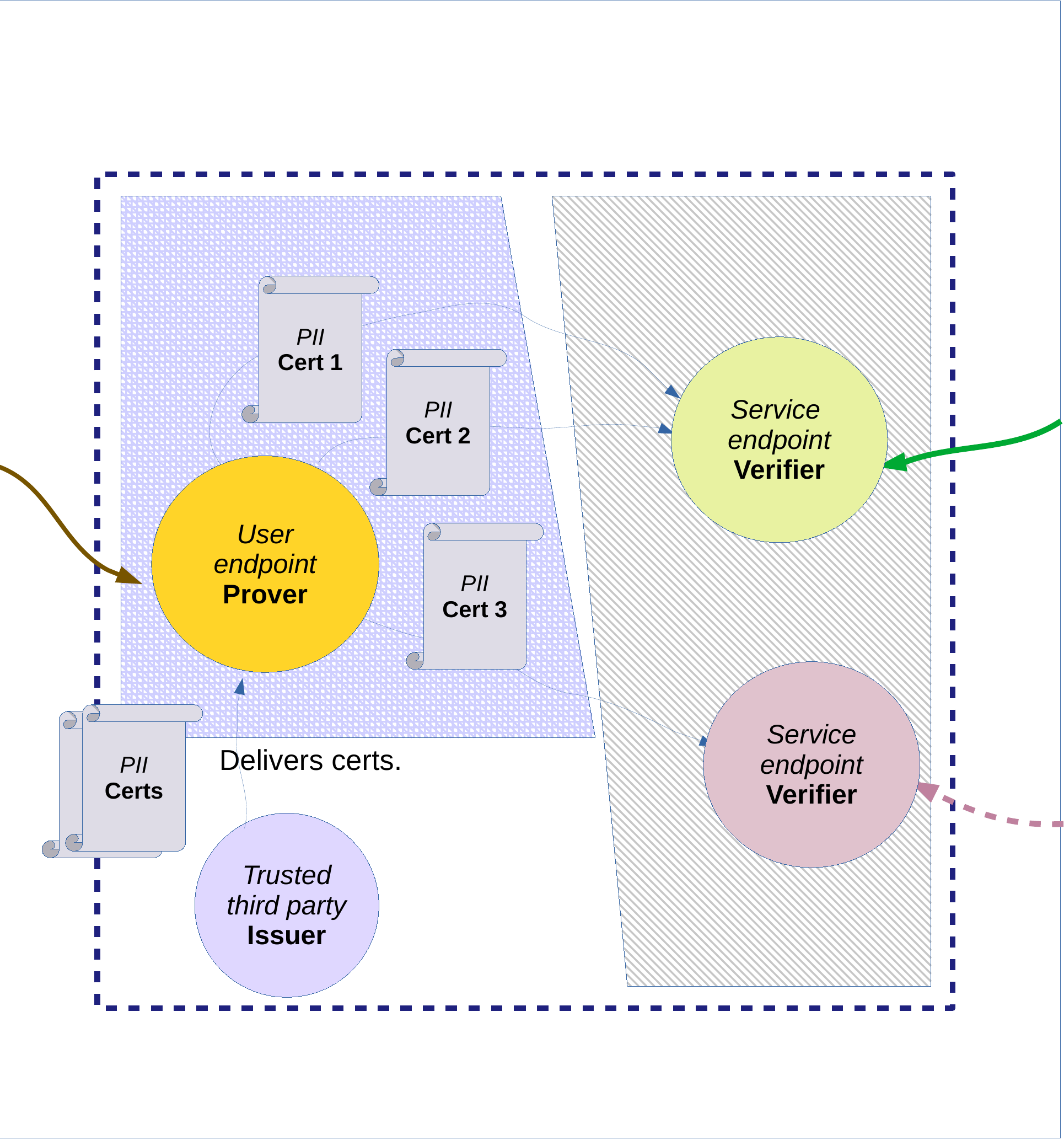}
\captionof{figure}{Anonymous certificate architectural layout diagram.\label{anoncred_arch}}

\subsection{Access-control delegation architectures: {\it (User-Managed Access, INDIGO)}\label{delegarch}}

These architectural solutions enable users to delegate access control to a dedicated software agent, acting as an authorization server.
This server deals with resource access on one or several resource server(s).
The resulting category studied here stands at an upper, more abstract level than the three other categories of solutions:
instead of proposing a single software tool, this category specifies the interactions happening between the different software entities ensuring PII management.

This type of solutions puts a strong emphasis over delegation, enabling the users to define how access control should be handled in an autonomous manner.

They enforce role decorelation over the system by keeping authorization and access control, on the one hand, and data storage, on the other hand, split into separate logical entities.
The SPs are then given data access through a standardized interface.

The goal of the decorelation is to ensure that the various responsibilities are dispatched evenly over the entities.
In case of one or more entities acting maliciously, this role decorelation process also reduces the risk of a total failure of the architecture.

Studying this category of solutions is assessing whether:
\begin{itemize}
\item a single software solution may not be enough in order to enforce our use case (Section~\ref{usecase}),
\item or an architectural pattern along with a communication protocol and specified interfaces should be chosen.
\end{itemize}

\textbf{\textit{Selected solutions}}

The two selected solutions for this category are User-Managed Access~\cite{UMASpecs} and INDIGO~\cite{CeccantiINDIGO}.

User-Managed Access is an OAuth2 profile meant for access control delegation.
It requires a five-party OAuth2 architecture, as depicted in Figure~\ref{deleg_arch}, in order for the user to enforce access control delegation, and for SPs to consistently access the user's PII.

As mentioned above, five entities contribute to this architecture.
The user appears as a resource owner (RO).
The user is also responsible for her/his PII management on one (or several) resource servers (RS), and for managing consent to SPs on the authorization server (AS).
The SPs appear as requesting parties (RqP).
The SPs interact with the authorization server, one or several (RS) and indirectly with the resource owner through a client (C).

The PII access happens in three steps:
\begin{itemize}
\item (RO) declares the protection upon a resource either before or while (C) attempts any access on it.
This configuration uses UMA's protection API, and is performed at resource registration;
\item (RqP) obtains authorization on (AS) through (C).
If the user consent has not been obtained before, it may obtained during that step (through the interactive claims gathering).
If the authorization succeeds, an access token (``Requesting Party Token'') is given;
\item (RqP) uses the access token on (RS), through (C), in order to obtain the resource.
\end{itemize}

User-Managed Access also specifies an offline decision algorithm based on the client's requested scopes in comparison with the client's previously granted scopes (\textit{e.g.}, when the user was online).
Currently being reviewed by the IETF for publication as a Request for Comments (RFC), User-Managed Access is meant to enhance the supported OAuth scenarios by bringing further delegation.
Although OAuth itself leaves room for delegation, many of its actual delegation technical- and implementation details are out of scope of the OAuth 2.0 official specification.

INDIGO, on the other hand, is a cloud computing software suite for authorization and authentication support and targeted at scientific communities.
It enables researchers to share documents and data using complex access-control enforcement scenarios.
Although meant for collaborative research efforts, it supports numerous features relevant to personal data self-management involving interactions with ASPs and PSPs -- mainly thanks to its emphasis on protocol interoperability, access control delegation, and users' official-identities management.

INDIGO's architecture theorizes solutions for a certain number of identity management problems.
For instance, protocol interoperability is ensured thanks to a token translation system.
Token translation means that the user can authenticate by using different schemes (SAML, OIDC, or plain X.509 certificates).
The SPs accessing the user PII then need to interface the solution as OIDC relying parties.
The token translation service also enables the non OIDC-compliant SPs to interface with the solution, using other identification protocols.

\includegraphics[width=\columnwidth]{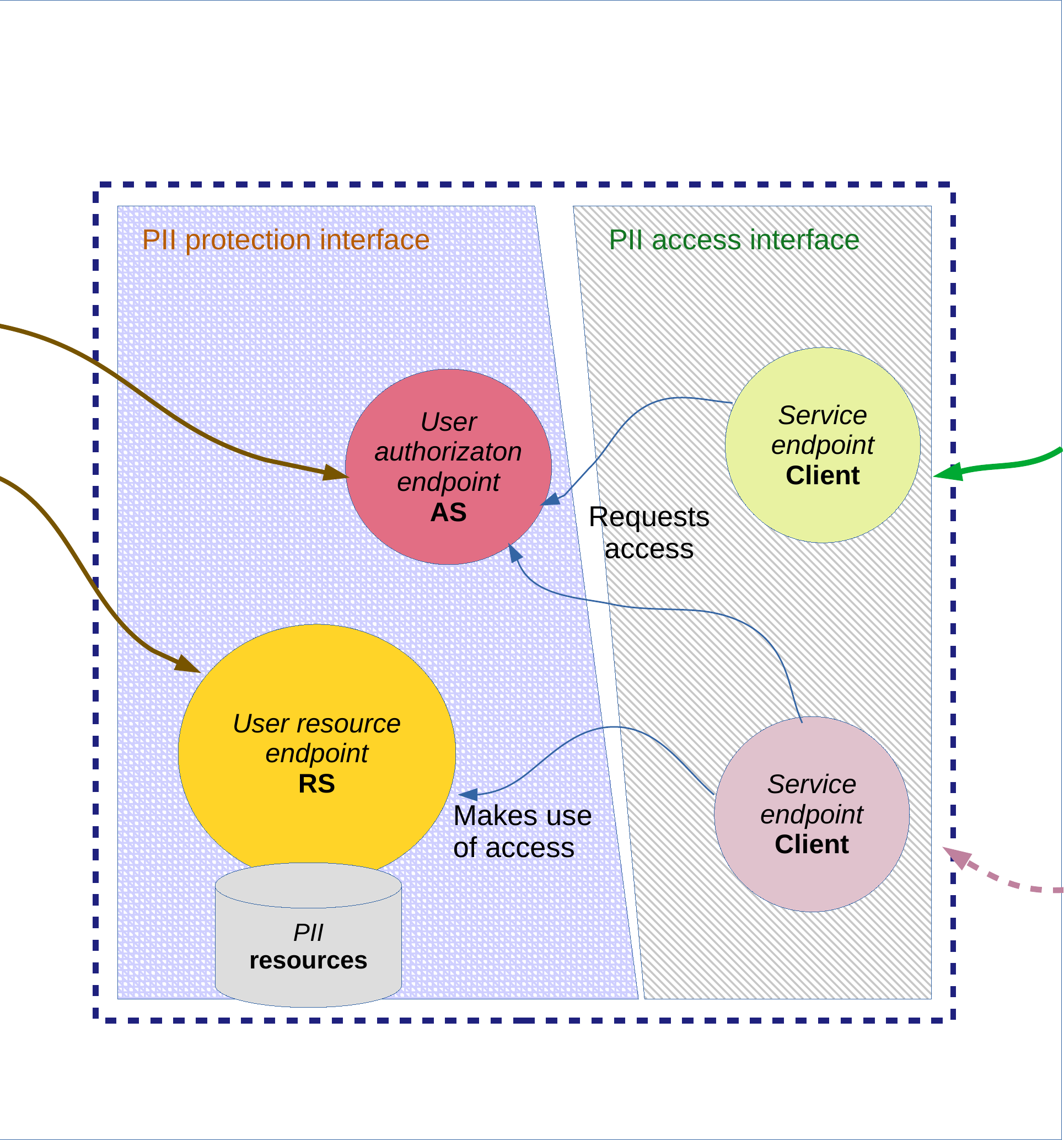}
\captionof{figure}{AC delegation architecture layout diagram.\label{deleg_arch}}

\section{Evaluation of the selected solutions\label{evaluation}}

This section provides a full and accurate analysis of the categories of solutions down to individual solutions, with regards to the eighteen criteria identified in Section~\ref{criteres}.

A synthetic comparative evaluation is given in Tables~\ref{synthtab} and~\ref{synthtab2}.

\setlength\tabcolsep{3pt}
\begin{table*}
\renewcommand{\arraystretch}{1.5}
{\tinysize
\caption{Comparative Evaluation of PII Self-Management Solutions for User Governance Criteria}
\begingroup
\setlength{\tabcolsep}{7pt}
\begin{tabularx}{17cm}{|X||X|X|X|X|X|X|X|X|X|X|}
\hline
\backslashbox[15.5mm]{Criterion}{Solution}
& BlindIDM
& Authentic
& Federation IdMs
& openPDS SA
& Mydex
& Databox
& Fargo
& Anon. Cert.
& UMA
& INDIGO\\

\hline

References
&   \cite{Nunez2014}
&   ---
&   ---
&   \cite{10.1371/journal.pone.0098790}
&   \cite{Papadopoulou2015}
&   \cite{DBLP:journals/corr/HaddadiHCCMM15}
&   ---
&   \cite{Camenisch2007,u-prove-technology-overview-v1-1-revision-2}
&   \cite{UMASpecs}
&   \cite{CeccantiINDIGO} \\

\hline
\hline

Type(s) of supported access-control
&  No info. avail.
&  (RB)AC on the IdP, for each federation
&  Multiple AC models on the IdP, for each federation
&  Simple grant/revocation of local PII processing requests
&  Temporal ACLs on PII or services
&  Global {\it Manager} with local delegation to services
&  No AC implemented
&  Not applicable
&  Double interface (protection API and token endpoint)
&  OAuth2 authorization model\\
\hline

Privacy usability trade-off
&  IdM can't read PII
&  None (Web identity federation configuration)
&  None (Web identity federation configuration)
&  Computa-tional trade-off (The PII processing has to be local)
&  User needs to set ACLs
&  Requires local drivers
&  Not supported
&  Potential anonymity revocation
&  User needs to define delegation
&  User needs to define delegation \\
\hline

User interface
&  No info. avail.
&  Web UI
&  Web UI mostly. If not, set of APIs for UI frontend.
&  Through mobile app.
&  Native client UI
&  No info. avail.
&  Web UI
&  Native cert. management. app.
&  No info. avail.
&  Web UI \\
\hline

{\bf Consent management}
&  Through login
&  Through login + ACL on PII attrs by the admin
&  Through login + ACL on PII attrs by the admin
&  ACL on {\it Safe Answers} for SPs by the user
&  ACL on PII for SPs by the user
&  ACL, TTL, anon. (enforced by the {\it arbiter})
&  Not supported
&  ZKPs generated by the user
&  Protection API consent flow
&  OAuth2 consent flow \\
\hline

Nego. of data-collection parameters
&  SAML federation. Less critical, due to opaque PII (reencryption proxy)
&  SAML or OIDC federation configuration
&  SAML or OIDC federation configuration
&  Not applicable (no collection as PII processing is local)
&  Nego. client offered to the user
&  No info. avail. (depends on the driver implementation)
&  Not supported
&  Yes, when reaching contractual agreement
&  Depends on the client implementation (OAuth2)
&  Depends on the client implementation (OAuth2) \\
\hline

Service provider revocation
&  Yes (SAML identity federation)
&  Yes (SAML or OIDC identity federation)
&  Yes (SAML or OIDC identity federation)
&  Yes (ACLs)
&  Yes (ACLs or TTL on authorization)
&  Yes (ACLs or TTL on PII)
&  Not supported
&  Not applicable (user-triggered atomic transactions)
&  Yes, by the user, enforced by the UMA AS
&  Yes, by the user, enforced by the OAuth2 AS \\
\hline

PII collection purpose definition
&  Not relevant for IdM
&  Not relevant for IdM
&  Not relevant for IdM
&  Not applicable (no PII collection)
&  No info. avail.
&  No info. avail.
&  Not supported
&  Yes, during contractual agreement
&  Possible (depends on the client implementation)
&  Possible (depends on the client implementation) \\
\hline

Inter-user PII sharing
&  Not relevant
&  Not relevant
&  Not relevant
&  No info. avail.
&  No info. avail.
&  No info. avail.
&  Not supported
&  No info. avail.
&  Possible (federated authz)
&  Possible (depends on the implementation) \\
\hline

\textbf{Online / offline mode(s)}
&  Not relevant for an IdM
&  Not relevant for an IdM
&  Both for IdPs features (w/ offline tokens)
&  Both
&  Both
&  Both
&  Not supported
&  Both
&  Both
&  Both \\
\hline

{\bf Extent of delegation}
&  Not supported
&  Not supported
&  Limited delegation scenarios
&  Simple PII transfer delegation
&  Simple PII transfer delegation
&  Queries on agent acting on behalf of the user
&  Not supported
&  Full delegation possible, depends on implementation
&  Complex delegation possible
&  Simple PII transfer delegation \\
\hline

History / logging of transfers
&  No info. avail.
&  Yes, for admin \& users
&  Yes, for admin at least
&  Audit log UI avail.
&  No info. avail.
&  1 data store dedicated to logging
&  For the platform admin only
&  No info. avail.
&  No info. avail.
&  No info. avail. \\
\hline

\end{tabularx}
\endgroup
\label{synthtab}
}
\end{table*}

\begin{table*}
\renewcommand{\arraystretch}{1.5}
{\tinysize
\caption{Comparative Evaluation of PII Self-Management Solutions for Data Exchange Flow Criteria}
\begingroup
\setlength{\tabcolsep}{7pt}
\begin{tabularx}{17cm}{|X||X|X|X|X|X|X|X|X|X|X|}
\hline
\backslashbox[15.5mm]{Criterion}{Solution}
& BlindIDM
& Authentic
& Federation IdMs
& openPDS SA
& Mydex
& Databox
& Fargo
& Anon. Cert.
& UMA
& INDIGO\\

\hline

References
&   \cite{Nunez2014}
&   ---
&   ---
&   \cite{10.1371/journal.pone.0098790}
&   \cite{Papadopoulou2015}
&   \cite{DBLP:journals/corr/HaddadiHCCMM15}
&   ---
&   \cite{Camenisch2007,u-prove-technology-overview-v1-1-revision-2}
&   \cite{UMASpecs}
&   \cite{CeccantiINDIGO} \\

\hline
\hline

Type(s) of supported PII
&  Data, attr. and value metadata
&  Data, attr. and value metadata
&  Data, attr. and value metadata
&  Metadata
&  Data
&  Data, doc.
&  Doc.
&  Data, doc., metadata
&  Data, doc., metadata (depends on the implementation)
&  Data, doc., metadata (depends on the implementation)\\
\hline

{\bf PII validation}
&  No info. avail.
&  Yes
&  No info. avail.
&  Not supported
&  No info. avail.
&  Not supported
&  Not supported
&  Yes
&  No info. avail.
&  No info. avail. \\
\hline

Functional structure
&  Web client, IdM server
&  Web client, IdM server
&  Web client, IdM server
&  Backend PII store, PDS frontend with SA modules, clients
&  Nego. agent, client
&  Primary and derived stores, drivers, central manager, apps
&  Web server with relational database backend
&  Issuer, prover, verifier, revocation referee
&  Five parties (RO, RS, AS, C, RqP)
&  IdM, AS and Token Translation Sys. \\
\hline

Prov. \& deprov. management
&  No info. avail.
&  By the user, or SPs through user management REST API
&  By the user, or SPs through user management REST API
&  Deprov. by the user or SPs
&  By the user or SPs
&  By the user or SPs
&  By the user or SPs
&  By the user (PII is obtained from the issuer). Deprov is less critical (pseudony-mity or anonymity, zero-knowledge)
&  User through RS, or SP after authz obtained from AS
&  User, or SP after authz obtained from AS \\
\hline

\textbf{Reusability of previously uploaded PII}
&  Yes, identity is federated
&  Yes, identity is federated
&  Yes, identity is federated
&  Yes
&  Yes
&  Yes
&  Yes
&  Yes
&  Yes
&  Yes \\
\hline

Minim. management
&  Minim. towards the IdM itself only
&  No
&  No
&  Yes, no raw metadata transfer
&  No info. avail.
&  No info. avail.
&  Not supported
&  Yes
&  Possible, client implementation dependent
&  Possible, client implementation dependent \\
\hline

{\bf Support of remote PII sources}
&  No info. avail.
&  Limited: support of user account backends
&  Limited: support of user account backends
&  Not supported
&  Not supported
&  Depends on driver implementation
&  Not supported
&  No info. avail., possible in future implementations
&  Possible, client impl. dependent + federated authz
&  Possible, client implementation dependent \\
\hline

\end{tabularx}
\endgroup
\label{synthtab2}
}
\end{table*}

The six critical criteria identified in Section~\ref{criteres} appear in bold font in these two tables.
For readability purpose and at no inaccuracy price, several solutions were gathered in the same column, as several approaches are close enough on a functional basis.
Thus, one column ``Anonymous Certificates'' gathers the two anonymous certificate solutions Idemix~\cite{Camenisch2007} and U-Prove~\cite{u-prove-technology-overview-v1-1-revision-2}, and column ``Federation IdMs'' gathers OpenIDM, Keystone and Keycloak solutions.

\subsection{Type(s) of supported access control}

The administrator of federated IdMs enforces the main configuration for data management, leaving the user little flexibility for adapting preferences to her/his own needs.
On the contrary, on the platform administrator side, such federation IdMs offer rich access-control configuration.
For instance, Keycloak combines different access control rules involving roles, user attributes and contextual authorization information.
Federated IdMs also cascade access control to service providers as part of the identity federation process\footnote{OpenIDM delegates the federation logic to another module, OpenAM. For readability purposes, and as the two modules are complementary, only OpenIDM will be mentioned in this survey. For more information regarding, OpenAM see \url{https://backstage.forgerock.com/docs/am} (last accessed: March 10, 2021).}.
Authentic whose access control model is much simpler than the two other federation IdMs selected for this survey, widely relies on the role-based access control model (RBAC~\cite{Osborn2007}) defined in the Django Web framework.\\
Each user is registered to a set of roles, and for each role, a set of permissions is granted.
At a given time the user's authorizations are deduced from the permissions of whole set of roles.
Some RBAC implementations also support role sessions, in which the activation/deactivation of a role for a given user may dynamically change over time.
A pillar concept of RBAC is the role hierarchy~\cite{Ferraiolo2001}, in which junior roles can be derived from senior roles according to the following transitivity rules:
\begin{itemize}
\item Permissions defined on junior roles transit up to senior roles.
\item Any user who is assigned senior roles also obtain their junior roles.
\end{itemize}
Whether the role hierarchy model supports the presence of cycles (in the graph of roles and their parenthood links) is implementation dependent.

Each PDS solution supports one or several specific access control mechanisms. \\
Mydex~\cite{Papadopoulou2015} displays an ACL configuration interface to the user. \\
Each rule in the list can be limited in time, assigned to a restricted subset of services, and revoked at any time. \\
openPDS~\cite{10.1371/journal.pone.0098790} proposes a simple grant and revocation mechanism, ensuring the legitimacy of SPs to send requests. \\
The access control in Databox~\cite{Mortier:2016:PDM:3010079.3010082} happens at two different stages: \\
$\bullet$ the local drivers offer an interface between local data stores and SPs.
Databox requires the deployment of one driver for each connected SP.
The access control performed at driver-level is made of lists (ACLs). \\
$\bullet$ the central Manager is the main management entity for the functional ecosystem.
It coordinates the access control on a more global scale.
This access control being performed by a unique and central authority is a mandatory access control system, as presented in \cite{DeCapitanidiVimercati2007}.
Access control is enforced at the Manager level through some access tokens.
Fargo does not implement any thorough access control.\\

Access control is out of the scope of the anonymous certificate use cases presented in \cite{u-prove-technology-overview-v1-1-revision-2} and \cite{Camenisch2007}. \\
More precisely, the user decides which  SP the transfer shall happen with.
The user is also able to manage her/his different anonymous certificates, possibly switching between multiple digital identities while interfering with different SPs. \\

UMA offers a twofold access control through its protection API and its access token endpoint.
INDIGO~\cite{CeccantiINDIGO} complies with the OAuth2 authorization model.
The access control is performed by an authorization server (AS), issuing access tokens and refresh tokens to SPs through the use of dedicated clients.
Theses SPs must have previously obtained the user's authorization.
The implementation of the dedicated clients should be OAuth2-compliant, and are not covered in the INDIGO presentation article.

\subsection{Privacy usability trade-off}
There is no visible trade-off for the user of federation IdMs and Authentic, as both the privacy and usability are enhanced by the federation process.
This absence of visible trade-off is made possible by identity federation protocols, responsible for PII transfer agreement directly between SPs and federation IdMs. \\
However, after the PII collection has happened, no FIM mechanism enforces that the PII is used according to the claimed purpose(s), as declared by the SPs. \\
Alternatively, BlindIDM~\cite{Nunez2014} supports more restrictive security hypotheses: it enforces PII privacy of the user from the IdM at no usability cost.
The reencryption proxy that BlindIDM manages prevents the IdM from reading the exchanged PII, as any other IdM could do, in a semi-honest manner.
Once the proxy has been established (more particularly, after the reencryption keys have been generated), this setting is transparent to the user.
However the number of keys managed by an IdM can increases rapidly: $n$ users connecting to $m$ SPs require the IdM to manage $n \times m$ reencryption keys. \\

openPDS, with its local computation process requires that the PII processing happen locally.
The solution therefore has a computational trade-off at ensuring user privacy: the PDS is responsible for the data processing, on behalf of the SPs.
There is also an SP-side of the usability trade-off: the SPs must send requests to the PDS using the specific SafeAnswers interface\footnote{This second type of usability trade-off could not be assessed in this survey, as no further information could be found about the SafeAnswers module.}.
Nonetheless, nothing in the related article asserts any trade-off from the point of view of the user. \\
Adopting a more conventional approach, Mydex requires that the user defines some access control lists for her/his PII.
It is considered as a usability trade-off from the point of view of the user, as some users may not want to define such ACLs themselves. \\
Databox relies on one local driver for each of its data store.
A trade-off from the point of view of the implementers, is obviously specific driver implementations.
As far as the article goes, there is however no visible trade-off perceived by the user, and there is no trade-off for the SP either. \\
Privacy enforcement mechanisms for Fargo are rudimentary, as a result there is no usability trade-off. \\

With anonymous certificate systems, the minimal disclosure of knowledge is under full control of the user, as explained in Section~\ref{anoncred}.
There is no user trade-off, as only the user is responsible for the way she/he manages her/his different anonymous certificates.
Nonetheless, the user should know that anonymity might be revoked. \\
The SP trade-off is the necessity to implement a zero-knowledge proof verification procedure. \\

Access control delegation architectures require that users explicitly grant authorizations at first.
The authorization grant must happen either before the PII transfer, of during it.

\subsection{User interface}
For federation IdMs, the FIM authentication process is transparent to the user.
The authentication scenario from the user point of view, whose simplified representation is in Figure~\ref{fig:fimflow}, does not change in spite of the separation between SPs and identity providers.
The authentication procedure itself is simple, usually relying on a simple login prompt on a Web HTML page.
The resulting SSO-authenticated session is also transparent to the user. \\
The identity federation process therefore does not impact the user interface, which stays as simple as it would be for a non-federated SP.
This is the case for federation IdMs as well as for Authentic. \\
As a result, and as a downside of this seamless interface, experience proved that the user of federated SPs does not always understand the federation process.
Additionally, when SPs require the collection of the user's PII, either the IdM acting as an identity provider or the SP must offer a consent collection page to the user. \\
BlindIDM~\cite{Nunez2014} gives no information about its user interface. \\
Eventually, Keystone, on the contrary, works as a set of identity management services whose only interfaces are APIs.

PDS solutions provide a limited informational user interface, whose main purpose is for the user to manage her/his PII, with optional applicative configuration capabilities. \\
openPDS, through its mobile app, proposes a UI for the configuration of the SafeAnswers module.
This interface outputs (i) the questions ``asked'' by the SPs to the PDS, (ii) the answer provided by the PDS, as well as (iii) the PII used for the computation of the answer.
The UI also makes it possible for the user to view the number of PII processing requests sent by an SP over a given period of time. \\
The UI proposed by Mydex allows the users to configure a set of ACLs ensuring that data collection is adequately configured, as shown by the visual elements in~\cite{Papadopoulou2015}.
As illustrated by these visual elements, the user is displayed the purpose of the PII collection, for one particular PII type.
They can define which of the four CRUD actions the SPs can take for this type of PII.
They can define a time-to-live value (TTL) for the PII retention.
Eventually, they can also decide whether these SPs are allowed to share this data with third-parties SPs.
However, the TTL defined by the user is enforced on the PDS only, and no mechanism is proposed for the enforcement of PII TTL when it has already been collected by SPs. \\
Fargo proposes a simple list of uploaded documents through its Web UI.
The user is able to perform simple operations such as uploading, downloading or deleting document. \\

The specifications provided for anonymous certificate systems do not cover the user interface.
U-Prove~\cite{u-prove-technology-overview-v1-1-revision-2} provides a native desktop client application.
The client providing the UI helps the user communicate with the certificate issuer.
It enables the user to understand the contractual agreements met with the issuer, including the anonymity revocation policies.
The interface enables the user to derive proofs of knowledge based on these certificates.
Eventually, it also helps the user communicate with SPs acting as verifiers. \\

Neither UMA nor INDIGO specifies any UI.
The software implementation has to design and provide a UI regardless of the specification documents for these two solutions.

\subsection{Consent management}
When using IdMs, consent is given by the user through a successful authentication phase.
For some identification protocols such as OIDC, the IdM acting as an identity provider shows the user a list of PII pending for her/his approval before collection by the service provider.
According to the European regulation, the consent given by the user is valid provided that the user understands the purposes of the PII collection. \\
Alternatively, BlindIDM also enforces ACL diffusion.
It uses signed cookies (``Macaroons''~\cite{41892}), which are suitable for carrying user consent information, however, without information about the possibility for users to trigger the generation of such signed cookies carrying caveats of their choice. \\

With PDS solutions, the approach is different.
As the SPs interacting with openPDS only get to receive the processed output for PII processing algorithms, the consent management for this solution consists in configuring how this local processing should happen.
The user is able to define which SP is able to send requests, and this definition stands for user consent. \\
Alternatively, Mydex's consent management simply consists in maintaining a list of AC rules, defined by the user, for each SP.
The user does not have any finer-grained consent mechanisms. \\
Databox lets the user consent to PII collection under some conditions.
For instance, the user may define a TTL function for some given PII, or choose an anonymization algorithm to be ran against the PII before any collection happens.
In spite of the rather decentralized architecture of Databox, these functionalities are all performed by the central arbiter. \\
Eventually, the rudimentary consent management capabilities of Fargo reflects its basic PII management features: the user, which manually manages the documents of the PDG through CRUD operations, also manually consents to any single reuse of these documents in an administrative online procedure.

With anonymous certificate systems, the user's consent is the decision to disclose the knowledge of some of her/his PII to SPs.
Thus data collection is initiated by the user and not the SPs. \\

Access control delegation solutions are meant to tackle user consent issues through a consent management entity. \\
In UMA, this entity is the authorization server, whose interactions with the SPs are performed using its access token and authorization endpoints.
The access policies are defined on the authorization server at resource registration, using its protection API. \\
INDIGO also implements an OAuth2 authorization server, but does not specify how the the implementation should manage user consent.

\subsection{Negotiation of data collection parameters\label{eval-nego}}
IdMs behaving as identity providers in a FIM system may adopt the negotiation mechanisms supported or suggested by identity management protocols (\textit{e.g.}, SAML and OIDC).
Users might be able to validate the list of their PII authorized to be sent to the SPs during the authentication step (cf. Figure~\ref{fig:fimflow}).
These pieces of PII are the identity attributes linked to the user's account (first name, last name, email address, phone number, \textit{etc.}). \\
However, this implementation-specific validation step often happens on an all-or-nothing basis and leaves no room for proper negotiation.
Typically, the list of PII collected by SPs is defined by the IdM administrator. \\

Mydex provides the user with a negotiation client, helping them meet a contractual agreement on PII transfer with the SPs.
In return, the SPs are given an agent allowing them to perform that negotiation step.
The negotiation client is then responsible for retrieving the PII processing policies for a given SP.
The client displays the policy parameters to the user, who in return can override them with their own preferences.
The negotiation agent, acting on behalf of the SP, decides which user requirements can be met, and which will have to be declined.
Eventually, the user, depending on the negotiation's agent response regarding her/his preferences, chooses whether to validate or to abort the PII collection. \\
Alternatively, openPDS makes it possible to avoid any raw PII collection.
Safe answers are processed locally, on the PDS, and only the result is sent to the SP.
This result information, sent to the SP, is of a lower dimensionality than the original input PII, making it more difficult for an intruder to infer identifiable information from it.
For instance, it prevents reidentifying partially-anonymised data through usual approaches of data-linkage from multiple sources~\cite{HENRIKSENBULMER20161184}. \\
Databox leaves the collection implementation-specific modalities to the decentralized drivers. \\

Anonymous certificate systems require the user to establish a contract with the SPs, so as to decide which proofs of known PIIs have to be delivered to them.
The negotiation is not covered in the solution, instead it is part of the implementation-specific contractual agreement. \\

The negotiation of such parameters in delegation architectures needs to happen before any SP accesses the user PII.
The OAuth2 protocol -- used in both the UMA and INDIGO solutions -- leaves negotiation concerns to the implementation of the client. \\
UMA suggests the implementation of an interactive claims gathering process, potentially prone to supporting a user-driven negotiation, however without any further implementation details.
Additionally, the UMA data-usage specification document~\cite{UMADataUsage2010} declares that the entities from UMA-compliant solutions are able to define a legal or contractual agreement defining the rights and responsibilities of each party.

\subsection{Service provider revocation}
Authenticated session management by IdMs involves the support of SP revocation.
Using the SAML protocol, the user can explicitly log out of her/his FIM session through Single Log-Out (SLO) services.
Technically, the IdM, acting as a SAML identity provider, along with the SPs exposing SLO SOAP services for direct communication -- \textit{i.e.}, not involving the User Agent.
However, the OIDC identification protocol does not support any such revocation capabilities not requiring any interaction with the user agent.\\

With PDS solutions, the enforcement of simple ACLs for each SP allows for a revocation process compatible with our use case.
Such a revocation process is supported by openPDS. \\
Mydex also allows the user to define implicit revocations triggered after a given timestamp:
the user's authorizations for SPs to access PII can be held valid for a specific time-window only.
When that time window is over, the resulting revocation happens automatically. \\
Databox also ensures implicit revocation, as  users can define a TTL value on their PII.
Thus, SPs having previously collected the users' PII are supposed to discard them when the TTL reaches zero.
However, no technical enforcement of this TTL mechanism is presented in the related article \cite{Mortier:2016:PDM:3010079.3010082}. \\

With anonymous certificate systems, certificate information is self-contained, and potentially meant for single-use only.
Hence the transaction involving the proof of knowledge is atomic and user-triggered.
As a result there is no SP revocation per se.\\

Finally, in delegation architectures, the central authorization server manages, on behalf of the user, the authorization grants and decides whether to renew access authorizations to SPs.
Optionally, and depending on the implementation, the authorization server may expose an access policy definition UI for the resource owner.

\subsection{PII collection purpose definition}
IdMs do not handle the concept of PII collection purpose definition.
Instead, it is the role of the SPs to declare the reason for the PII collection.

Thus the user is asked for her/his consent for a particular purpose.
However, for this category, no mechanism enforces that SPs use the collected PII according to the declared purpose(s).
As a result, this criterion is not relevant for this category of solutions. \\

Among the PDS solutions, only openPDS addresses this concern, however indirectly:
SafeAnswers ensures that the PII processing happens locally.
Therefore, rather than enforcing collection purposes definition for SPs, openPDS makes it possible to view the underlying PII processing algorithm. \\

The purpose of the PII certificate issuance is defined when reaching the contractual agreement between the user and the certificate issuer. \\

Eventually, UMA and INDIGO do not directly consider this criterion, which is left to the client implementation.
For legal reasons, the client implementation has to deal with PII collection purpose definition concerns.

\subsection{Inter-user PII sharing}
IdMs are meant to help the user deal with her/his own digital identities, and user-to-user PII sharing are not part of their concerns.
This criterion is not relevant for this category of solutions. \\

As for PDS solutions, the articles presenting Mydex, openPDS, and Databox do not mention whether such a feature is directly possible thanks to these three solutions. \\
We note that many of the use cases presented in the related articles deal with e-health concerns.
These concerns are not directly compatible with inter-user PII sharing features. \\

Also, use-cases including data sharing between different users of anonymous credential systems have not been addressed in related articles. \\

The federated UMA authorization specifications~\cite{Kantara2018UmaFim} defines how resources servers belonging to different domains can be federated by a single authorization server, potentially enabling inter-user PII sharing. \\
INDIGO does not directly consider this criterion, which would depend on the implementation of a specific user client.

\subsection{Online/offline modes}
The various processes of identity management, \textit{e.g.}, identity provisioning in a FIM system, require the support of an online mode by any of the IdM solutions.
However, the offline mode is not strictly required.
Nonetheless, in spite of their primary purpose of synchronous identity management and providing, some of the selected solutions offer an offline mode.
Federation IdMs as well as Authentic comply with the OpenID Connect specifications when acting as identity providers, and are able to deliver offline tokens to the OIDC relying parties.
These tokens may be, under certain conditions (especially their expiration timestamp), stored for later use by the relying parties.
Still according to the OIDC specifications, the federations IdMs can deliver refresh tokens to their relying parties, enabling later offline access rights renewal. \\

The support of such modes varies from one PDS solution to another.
First, openPDS supports an offline mode, as the users are able to define how the PDS should act on their behalf, dealing with SP requests while they are offline. \\
Databox also offers such an offline mode: local PII store drivers also act on behalf of the users. \\
Eventually, Mydex lets the users define access control rules that apply even when they are offline. \\

Idemix~\cite{Camenisch2007} and U-Prove support both modes.
Apart from their ability to present standard (interactive) proofs of knowledge, an anonymous certificate system implementation can support non-interactive proofs, as originally suggested by~\cite{Rackoff1992}, which are perfectly suitable for handling the offline mode:
the non-interactive proofs can be computed first by the prover, and only then showed to a verifier, in an offline manner. \\

The approaches used by delegation architectures are designed to support both online and offline modes. \\
The authorization granted to the parties requesting access to PII on the resource server is not necessarily synchronous.
Depending on the authorization server configuration, the client, acting on behalf of an SP, can obtain an access token, usually with a limited time-to-live.
A pseudo-algorithm, meant as an implementation guideline and visible in \cite[chap 3.3.4]{UMASpecs}, defines the conditions under which the authorization server delivers an access token, and which scopes are granted with it.
Granting scopes in a delegated manner requires applying a short algorithm that involves the following elements:
\begin{enumerate}
\item The client's pre-registered scopes (\textit{i.e.}, the scopes registered before the authorization request).
\item The client's most recently requested scopes at the authorization server's token endpoint.
\item The scopes associated with the resources whose access is requested by the client.
\end{enumerate}
Additionally, UMA and INDIGO are based on the OAuth2 protocol, which is compatible with the enforcement of access control rules that apply even if the user is offline.

\subsection{Extent of delegation}
With IdM solutions, the user does not delegate any PII management task.
Some secondary modes such as offline token issuance imply, to some extent, user delegation -- however these secondary modes do not handle delegation as defined in the selected use case in Section~\ref{usecase}.
Keycloak and OpenIDM also serve as Authorization Servers in the User-Managed Access profile for OAuth2, which is the one more step to enforcing the required level of delegation.\\

The delegation capabilities of openPDS and Mydex are simple data access grants and revocations, for each SP. \\
Databox implements an agent acting on behalf of the user, after obtaining their access-control preferences regarding each SP. \\

Neither Idemix or U-Prove mention the extent of consent for delegation.
However, the certificate disclosure algorithm presented in \cite[Chapter~15.6.4]{Camenisch2007} (see the \texttt{ShowCert} primitive), revealing an anonymous certificate to an SP acting as a verifier, can be handled, depending on the implementation, by a software agent acting on behalf of the user. \\

Delegation architectures are by nature designed to implement authorization delegation.
The authorization server is the central delegation entity of the user, it enables the fine-grained consent scenarios that are necessary for our use case to be enforced.
It acts on behalf of the user when deciding whether to grant, to deny or to renew access to PII for an SP.
Usually, the user can define the access policies that apply at resource registration, on the authorization server.

\subsection{History/logging of transfers}
According to its configuration, the Web server bound to any Web IdM solution can log the transfer metadata from HTTP requests and responses (see the SAML~\cite{RFC7522} and the OpenID Connect~\cite{OIDCCore} protocol specifications for details about metadata format for Web FIM).
Additionally, the IdM may log FIM authentication sessions across service providers, for traceability or debugging purposes. \\

The ability to address this criterion differs from one PDS solution to another. \\
openPDS offers a log page to the user, giving her/his global information about requests sent by the SPs.
Fine-grained PII logging capabilities as required by our use case have not been presented in the related article~\cite{10.1371/journal.pone.0098790}. \\
Databox, on the contrary, deploys a specific data store whose role is to log any PII transfer.
However, the high-level presentation of the solution in the related article does not mention the exact content of the logs. \\
Fargo, as a Web application, stores HTTP metadata through its underlying Web server (either emitted or received).
Its rather basic supported use cases do not require fine-grained logging of PII transfers.
As a result logging information in Fargo is not made available to the user. \\

A logging facility is necessary to anonymous certificate systems if willing to support a revocation procedure.
The anonymous transactions may be logged so as to be retrieved in case of a contractual conflict between the prover and the verifier. \\

Eventually, neither UMA or INDIGO mention how the logging of transactions should happen.
However, the Kantara Initiative, editor of the  UMA specifications, has also proposed a consent receipt model~\cite{KantaraCRSpecs,KantaraMVCR}, which can be used for maintaining an history of the consents granted by the users.

\subsection{Type(s) of supported PII}
IdMs handle data, and attribute and value metadata linked to a user account.
This PII is transferred to SPs during user authentication phases. \\

Once again, each of the selected PDS solutions adopts a different approach. \\
openPDS is designed to perform PII metadata storage.
The use cases given as examples in \cite{10.1371/journal.pone.0098790} target biometrical PII metadata. \\
Mydex stores simple PII data. \\
Databox can store both simple PII data and user documents (see examples given in \cite{DBLP:journals/corr/HaddadiHCCMM15}). \\
Fargo offers simple user document storage features. \\

Proofs of knowledge are inherent to anonymous certificate systems, and can support any of the three types of PII discussed in Subsection~\ref{def-pii-types}.
These solutions allow the user to prove properties such as documents ownership, metadata validity and PII data authenticity. \\

At last, in access-control delegation architectures, the set of supported types are implementation-dependent.
As a result, neither UMA or INDIGO explicitly mention the list of supported types of PII.

\subsection{PII validation}
PII validation is not a core feature of IdM solutions.
However, Authentic provides rudimentary identity-attributes validation, meant for territorial authorities to validate attributes supplied either by the user or by ASPs. \\

The PDS solutions do not address the concern of data validation. \\
Mydex raises the concern of PII validation (see \cite[Chapter~4]{Papadopoulou2015}), however with no technical mechanism proposed to handle this process. \\

Anonymous certificate systems are designed for PII validation by a trusted authority (the certificate issuer). \\

Finally, PII validation concerns do not appear in the solutions' respective presentation articles.

\subsection{Functional structure}
Despite minor variations, the functional structure is the same for all the IdM solutions.
Keystone, Keycloak, OpenIDM, BlindIDM and Authentic all implement a federated-identity provider, relying on standards FIM protocols. \\

openPDS proposes a three-part architecture:
\begin{itemize}
\item the backend, hosting PII store(s).
\item the frontend, running the SafeAnswers module.
\item the client(s), \textit{i.e.}, the user-side parties of the solution.
\end{itemize}
Separate data stores are used in Databox.
A store is assigned a dedicated driver, and is executed within a Docker container environment.
These stores can be primary stores, containing raw PII, or derived stores, containing the results of PII processing.
The drivers take care of implementation-specific variations for each store.
Deporting the storage of data to a different -- and potentially remote -- component may be necessary in some scenarios (\textit{e.g.}, the deployment of data in cloud computing systems), but may also raise some particular security concerns.
Also, a central manager orchestrates the interactions between inner components, as well as the PII transfers with SPs.
All the stores offer the same exact API to the manager.
For security reasons, the different components are deployed in the same virtual private network (VPN).
Proxying to Web services is performed by a SOCKS server~\cite{RFC1928}.
Finally, third-party service providers implement applications performing PII management operations on the (primary or derived) available stores. \\
Mydex, apart from proposing a PII usage configuration client to the user, also offers a negotiation agent.
No information about its storage backend is given. \\
Fargo is a monolithic Web application, accessed from a Web browser by the user.
Its PII is stored on a relational database. \\

The functional structure of anonymous certificate systems is a four-party architecture as described in Section~\ref{anoncred}. \\

In the five-party UMA architecture, as presented in Subsection~\ref{delegarch}, the Client interacts with the user on behalf of requesting parties (either ASPs and PSPs).
Indeed, for any such party, the Client accesses authorization data on the authorization server, and eventually PII on the resource server(s). \\
A noteworthy property of this architecture is the decorelation that lies between two different interfaces: the protection API and the OAuth access token endpoint.
This separation is justified: these two interfaces fulfill two distinct roles, \textit{i.e.}, respectively letting the user manage her/his resources, and letting the SPs access PII depending on the user-defined authorizations on the AS. \\
The INDIGO architecture implements the OAuth2 framework, in which a central entity acts as an authorization server.
The overall functional structure is simpler than the one adopted by UMA.
However, INDIGO's Token Translation System (TTS) supports different identity management protocols, and enables interoperability between Web-based and non-Web-based services.
This TTS is also modular: the introduction of new types of tokens involves implementing new plugins.

\subsection{Provisioning and deprovisioning management}
Depending on the implementation, IdM solutions can offer an interface for the SPs to perform CRUD operations on user accounts. \\
Authentic and the three federation IdMs provide a user management REST~\cite{FieldingRest2008} API, suitable for performing both provisioning and deprovisioning operations. \\
They also support backends in order to retrieve PII from remote sources, \textit{e.g.}, user accounts from a remote directory server.

openPDS specifies a service protocol designed for read-only operations.
This solution therefore focuses on PII collection by services, not on its provisioning.
The PII is provisioned on the solution beforehand, through APIs.
No further technical details about Databox's provisioning capabilities are given in the presentation article. \\
Mydex supports data provisioning by a set of registered services.
The data flows happen bidirectionally, and services are able to collect or to provision data on the PDS.
The user's consent is defined using access control lists, regulating access to data for each registered service, for a given time window. \\
Databox also provides an interface for SPs to provision PII.
The access modalities of the interface is implementation-specific, and is not covered in the Databox presentation article. \\
Fargo offers provisioning and deprovisioning by ASPs and PSPs.
openPDS and Mydex also support PII deprovisioning by SPs. \\

For anonymous certificate solutions, the user is provisioned with PII certificates, at her/his own will, after contacting the issuer.\\
The anonymous certificate category does not support ``provisioning'', as defined in Subsection~\ref{crit-provdeprov}, as it does not enable either the ASPs or the PSPs to write or modify PII on the user-centric solution.
The only information revealed is the proof of knowledge of the user's PII. \\
Additionally, when the certificate issuer can be trusted -- which is the normal use case of the solution -- and when no anonymity revocation is happening -- meaning that there is not any contractual conflict between the different actors --, no unnecessary PII may be revealed to the SPs.
As a consequence, in case of a normal use of the system by all its actors, this category of solutions reduces the criticity of PII deprovisioning concerns. \\

UMA and INDIGO both rely on the OAuth2 protocol~\cite{RFC6749}.
They are able to support PII provisioning or deprovisioning, as this protocol allows for PII management operations on the resource server.

\subsection{Reusability of previously uploaded PII}
For IdMs solutions, the user PII stored as a profile account is by nature meant to be reused for later transfers. \\

The PII data collected by PDS is also meant to be reused.
Reusability is a core feature of this category of solutions.\\

The reusability of anonymous certificates is supported by this category of solutions.
This category of solutions also supports disposable certificates, although not relevant for our use case (see Subsection~\ref{usecase}).\\

Eventually, access control and authorization delegation also ensures PII reusability in a privacy-compliant manner.

\subsection{Minimization management}
The IdMs are not able to check whether PIIs requested by SPs are limited to what is strictly needed for providing services to user.
FIM protocols such as OIDC claims, enabling SPs to request particular pieces of PII according to a given profile.
However, no mechanism enables the IdM to verify that the claimed PII is truly needed for the actual services provided to users. \\

The PDS solutions implementing authorization protocols such as OAuth may propose minimization features.
Depending on the client implementation, the user may be able to view the list of data required by SPs and to deny the authorization grant if the list is not minimized enough. \\
openPDS performs PII minimization by providing only some ``safe answers'' to the SPs (see the explanation given in Subsection~\ref{eval-nego}).\\

Anonymous credentials systems are designed for data minimization support.
ASPs and PSPs only know the minimum proofs of knowledge, with the optional ability to reidentify users in case of conflicts. \\

Eventually, implementation-specific data minimization might be enforced with access-control delegation architecture solutions, at the client-side of the architecture.

\subsection{Support of remote PII sources}
Feration IdMs and Authentic support authentication backends:
they may be connected to remote sources such as directory servers which as a result can synchronize or duplicate partial or complete user account information.
This feature is however a strict subset of the functional expectations when applying this criterion to our use case.\\

For PDS solutions, only Databox provides a support of PII sources thanks to its extensible data-flow model.
However, this support depends entirely on the implementation on the driver for each remote source. \\

The support of remote PII sources is not covered in the articles presenting this category of solutions.\\

UMA supports federated authorization~\cite{Kantara2018UmaFim}, enabling remote OAuth2 resource servers (RS) to interact with a single authorization server.
Alternatively, non-RS remote sources can implement a client-side party, obeying to the supported client-authorization protocols of these solutions. \\

\section{Synthesis\label{synthese}}

This section provides a synthesis of the per-category evaluation presented in Section~\ref{evaluation}.
Beginning with global observations, it also presents specificities of each selected solution.

A brief analysis of how each category may be suitable for supporting administrative services -- determining whether these categories address the six critical criteria for our use case -- is also conducted.

Alternatively, a concise interpretation of the synthesis is provided by Table~\ref{summary_table}, directly linking the four categories and their inherent ability to address the critical criteria.

\setlength\tabcolsep{3pt}
\begin{table*}
\renewcommand{\arraystretch}{1.5}
{\tinysize
\caption{Summary of the Inherent Capabilities of Categories of Solutions to Address the Identified Critical Criteria}
\begingroup
\setlength{\tabcolsep}{7pt}
\begin{tabularx}{17cm}{|X||X|X|X|X|}
\hline

\backslashbox[34mm]{Criterion}{Solution}
& Identity Managers
& Personal Data Stores
& Anonymous certificate systems
& Delegation architectures\\
\hline

Consent management
& Supportable
& Supportable
& No need for it, minimization of collection
& Supported, by nature\\
\hline

Online / offline mode(s)
& Not relevant by nature
& Supportable
& Supportable (offline certs.)
& Inherently supportable\\
\hline

Extent of delegation
& Significant
& Implementation-dependent
& Not relevant
& Significant, by nature\\
\hline

PII validation
& Supportable, implementation-dependent
& Supportable, implementation-dependent
& Inherently supportable
& Supportable, implementation-dependent\\
\hline

Reusability of previously uploaded PII
& Limited, by nature
& Inherently supported
& Inherently supported
& Inherently supported\\
\hline

Support of remote PII sources
& Implementation-dependent
& Implementation-dependent
& Not relevant
& Implementation-dependent\\
\hline

\end{tabularx}
\endgroup
\label{summary_table}
}
\end{table*}

\subsection{Identity managers}

Identity managers provide PII exchange capabilities for their users.
The exchanges happen as part of identity management and identification protocols.
The Web IdMs require no additional component installation on the user's system -- apart from a standard Web browser.

However, most of the user-driven PII management features offered are not standardized, and they differ from an IdM solution to another.

Also, authorization protocols support several modes, profiles, grant types and authorization flows, thus leading to many varying implementations.
Even implementations of the same protocol can decide to adopt mutually exclusive subsets of these variations.
For instance, OpenID Connect specifies various profiles and multiple authorization schemes.

Moreover, these IdMs solutions also rely on their sets of specific security hypotheses.
For instance, Authentic is at the heart of the chain of trust.
By design, it manages the user's PII and has direct access to it.
Many user attributes and metadata, obeying to an extensible user attribute model, are stored in the IdM's database.
Although the FIM protocols it uses provide security and privacy properties even in case of untrusted service providers, a security failure within the IdM would result in user privacy threats.
On the contrary, BlindIDM supports much more restrictive security hypotheses, preventing the IdM to read the users' cleartext PII.

The evaluation of IdMs according to the five critical criteria reveals that:
\begin{itemize}
\item The support of remote PII sources is limited for some solutions of this category.
\item The consent management model remains implementation-dependent.
\item Simple delegation scenarios are possible only for a subset of solutions from this category.
No further delegation model is proposed.
\item The support of validated PII is not a core feature for this category of solutions.
\end{itemize}

More generally, IdMs do not provide all the necessary features required for the enforcement of our territorial use case.
From the user's point of view, IdMs enable identification and account self-management.
The other PII management features expected in this survey are by nature not applicable to IdMs.

Finally, targeted administrative services do not necessarily imply authentication nor they need to manipulate user accounts.
For instance, they may require only atomic PII transfers in order to fulfill a user request -- or the service can be used in an anonymous manner.
Eventually, mechanisms such as tracking codes enable users to view the processing of their requests, without imposing any account nor identification process -- hence without the need for identity management.

\subsection{Personal Data Stores}

Two different approaches in terms of user governance are retained by PDSs.

The first approach is the deployment of an online PII-storage instance common to several users.
As a result the storage entity is not a personal instance and is not entirely user-driven.
It may be deployed in cloud architectures, as does Mydex, designed specifically for this approach.
Such solutions into production usually refer to a business model such as Software as a Service (SaaS)~\cite{Turner2003}.

The second approach is a fully user-driven storage instance, as supported in Databox for instance.
It enforces user governance thanks to the physical ownership of the user's data.
This approach is particularly suitable for deployment on a user device (\textit{e.g.}, a smartphone or even specific SoC hardware).
In this scenario, the user has the complete responsibilities regarding the management and the transfer of the PII stored on the PDS.
This scenario is relevant for the respect of the user's privacy, provided that the user understands the PII management features offered by the solution.

As a consequence, PDSs offer a PII dashboard for users to manage their PII and optionally their previously given consent to PII collection.
The user's consent for data collection by third parties is made on an ``all-or-nothing'' basis: either users keep their PII private, or they make this PII completely available to a third party.
Additionally, these dashboard tools may enable users to visualize which services are registered.
They also offer data consumption parameters or metrics.
Optionally, they may even help users manage storage capabilities on the PDS.
The full PII lifecycle is therefore handled by these solutions.

For a given SP and a given PII processing purpose, the PDS must also obtain the user's consent before transferring any PII.

Eventually, evaluating PDSs according to our critical criteria reveals that:
\begin{itemize}
\item Databox indirectly addresses the support of remote PII sources.
\item Managing consent and delegation remains simple ; advanced scenario such as negotiation or partially-autonomous decision making are not supported.
\item However applicable by this category, no selected solution offers PII validation features.
\end{itemize}
Hence, for simple PII management scenarios, PDS are suitable for handling services offered by the administration and collectivities -- especially when these services require a recurrent collection of reusable PII.

\subsection{Anonymous certificates\label{synth-acred}}

Both anonymous certificate solutions Idemix and U-Prove implement the same set of PII management features.
They provide an elegant way to apply the principle of data minimization, as required by the current legislation in the European Union.
This principle is enforced by the minimal disclosure of PII.

These solutions are user-centric by design, but not entirely user-driven:
actions on the system such as certificate issuance, anonymity revocation and certificate verification are not led by the user.
Indeed, their trust model requires that a certificate issuer and a revocation referee be managed by trusted authority entities.
This model is suitable to the administration context.

Eventually, the revocation referee may act at the expense of the user.
The anonymity-revocation policies, as mentioned in the Idemix presentation article~\cite{Camenisch2007}, may happen without consent from the user.
This can happen in case of certificate misuse by the latter, \textit{e.g.}, when the user does not respect the contractual clauses defined with the SP.

Anonymous certificate systems provide an answer the critical criteria identified in Section~\ref{criteres}, as detailed below:
\begin{itemize}
\item Supporting remote PII sources is implementation-dependent, and is not addressed directly in the two articles presenting the selected solutions.
\item Delegation is possible, but also entirely depends on the implementation.
It is not covered in the two articles presenting this category of solutions.
\item PII validation is a core concept of this category of solutions.
\end{itemize}

As a result, anonymous certificate systems are suitable for some needs of the administrations and territorial collectivities.
They would prove to be useful when, instead of PII collection, the services only need the assurance that the user is legitimate, or that they detain the adequate information to use the services.
For instance, an ASP willing to compute some statistics regarding its users, without any involved PII but the assurance that the user is legitimate, could deploy an anonymous certificate system to obtain this legitimacy information from its users.

\subsection{Access control delegation architectures}

UMA stands out as it asynchronously delegates PII access control to a dedicated entity (an authorization server)
It provides two standardized interfaces on user side and SP side.
It decorelates the access policy definition, performed by the user, and the enforcement of these policies, performed by the authorization server.

INDIGO adopts a different approach.
Its token translation system makes it possible to support various authorization protocols.
It also enables the user to define a set of authorization rules, meant for a software agent to act on her/his behalf.
Although designed as a collaborative academic research tool, it satisfies several needs of our selected use case.

The critical criteria are addressed by this category of solutions as below:
\begin{itemize}
\item The support of remote PII sources is implementation-dependent.
\item Delegation and consent management are core features.
\item PII validation is not covered in the related literature.
\end{itemize}

More generally, this category of solutions is suitable for services offered by administrations and collectivities in which repeated authorization decisions on behalf of an offline user have to be made.
Eventually, the user may appreciate being able to define a set of authorization policies, which are then enforced by the solution.

\subsection{Identifying an optimal solution}

In order to meet our use case, we can provide layout for a solution matching as many criteria as possible, thus making this solution optimal.
In particular, this optimal solution has to enforce the five critical criteria identified in Section~\ref{criteres}.
As per our use case in Subsection~\ref{usecase}, deploying an access management ecosystem -- \textit{i.e.}, the second family of solutions according to the taxonomy performed in Figure~\ref{arbre} -- is not necessary and can't cover our five critical criteria adequately.
In particular, the properties enforced by this category of solutions are at too high an abstraction level and leave too much to potential implementations of subcomponents of such solutions for them to be considered optimal.
Moreover, as explained in Subsection~\ref{synth-acred}, anonymous certificate systems are relevant in only a subpart of our use case.

The five critical criteria can be enforced with a solution belonging to the family of single software entities according to the classification performed in Figure~\ref{arbre}.
Identity providing may be necessary, but it is not central enough for the use case for an IdM solution to be selected as the optimal solution.
Additionally, the needs for the optimal solution to support remote sources, extended delegation schemes and both online and offline modes disqualifies IdMs.

Instead, we identify the need for an augmented PDS tool, which would provide the following features:

\begin{itemize}
\item An extensible remote-source support model.
Indeed, the administration and collectivities already offer a wide variety of PII sources, made available to any of their authenticated users.
These sources answer to the public services digitization initiatives amongst the administrations and collectivities of European countries.
For instance, the French authorities maintain their official APIs\footnote{See \url{https://api.gouv.fr/} (last accessed: March 10, 2021).}.

This feature directly addresses the criterion described in Subsection~\ref{crit-remote-sources}.
Among PDS solutions, Databox proposes a modular and partially decentralized  approach, which can be extended in order to fully support remote PII sources (1).
Enhancing this model would make it possible for a PDS to abstract the PII location, \textit{i.e.}, to support it whether it is locally hosted or remotely available.

\item A clear interface of PII consumption directly mapped to the user's previously given consents.
This feature addresses the criterion described in Subsection~\ref{crit-consent-management}.
The support of consent metadata within PDSs has already been widely discussed, which eases the research work in order to implement proper consent metadata within a PDS.
The optimal solution could for instance implement the management of consent receipt as defined by the Kantara initiative~\cite{KantaraCRSpecs}.
A optimal solution supporting such consent receipts would benefit the users as well as service providers, for both functional and legal purposes.
Using this consent model, PDS as a monolithic logical entity could bear the roles of resource- and authorization-management.
Conformance to authorization management protocols such as OAuth2 is a first step towards this objective.
\item The ability to validate PII for simpler user-relationship management processes.
This addresses the criterion described in Subsection~\ref{crit-pii-validation}.
Such PII validation, implemented for instance by adding a simple boolean flag value to each element of the PII data model in the solution, would lead to simplifications in several PII management processes.
\item The ability to act on behalf of the users for the multifold steps of PII management -- from the creation or collection of this PII on the platform, to its use with ASPs and PSPs and eventually its deletion.
This addresses criterion described in Subsection~\ref{crit-extent-delegation}.
\item The ability to ensure PII management when the user is not connected to the platform, which is a direct consequence of the conformance to the critical criterion described in Subsection~\ref{on-off}.
This requirement is associated with the ability to reuse previously uploaded PII -- addressed by the critical criterion described in Subsection~\ref{reusability}.
\end{itemize}
The support of remote PII sources also brings some more specific problems such as the automated matching of identity attributes retrieved across several sources.

As a result, a PII store server acting as a PII source hub should be chosen, minimizing the cumulated efforts necessary for a research contribution in order to handle all the aforementioned features.

\section{Future research directions\label{research-directions}}
The future research directions for the PII source hub mentioned in Section~\ref{synthese} are as follows:
\begin{enumerate}
\item The definition of a consent model that would enable the support of the relevant critical criteria, \textit{i.e.}, the support of PII sources, the extent of delegation \& consent management, and the support of online \& offline modes.
Indeed, solutions managing a thorough consent model relevant to the territorial use case, such as~\cite{KantaraCRSpecs} have a tremedous advantage for enforcing subparts of the use case.
They provide better traceability and enable delegation capabilities.
\item The interoperability concerns that arise when dealing with PII sources of different types.
For instance, some sources may support plain HTTP Basic authentication~\cite{RFC7617} while others may implement the OAuth 2.0 authorization framework~\cite{RFC6749} instead.
The implementers of such a PII source hub may want to support the OAuth 2.0 assertion framework as specified in~\cite{RFC7523} in order to provide further interoperability, for instance by being able to interface with SAML-based sources~\cite{RFC7522}.
\item The possible matching issues when dealing with several PII sources.
That is, pieces PII prodived by different sources regarding a user should be matched to ensure they describe the same actual user.
Failing to perform identity matching enables malicious entities to perform attacks such as user collusion, leading to applicative privilege escalation.
The underlying protocols of resource access management do not provide ``out-of-the-box'' solutions in order to perform identity matching.
In order to prevent such types of attacks, such a PII source hub should be able to match pieces of PII according to their redundancy over different PII sources.
The algorithm used to detect non-matching PII should be able to identify ambiguous cases which require the interaction of the territorial collectivity (human) agent.
\end{enumerate}

\section{Conclusion\label{conclusions}}

This paper surveys the technologies addressing personal data self-management in the context of administrative and territorial public service providers.
The resulting comprehensive and comparative study identifies the current limits of these technologies, specifically with regard to our six identified critical criteria.
We observe generally that there is no definitive position of the approaches against these six criteria, and that some of them are designed to address one of the criteria, but not all six at once.

Eventually, the scientific literature such as \cite{Papadopoulou2015} and \cite{Andersdotter2016} proves that many territorial collectivities are willing to participate in pilot projects regarding new PII management solutions.
The administration and collectivities wish to enforce their status of authoritative entities providing digital services, and understand the need to be flawless regarding adequate PII protection for the users of SPs they offer, including by testing innovative solutions.
These pilot projects are also a way to raise users' awareness, who in most cases, are not aware enough about their rights to privacy.

\end{multicols}

\begin{multicols}{2}

\noindent\parbox{8.3cm}{\parpic{\includegraphics[scale=.15]{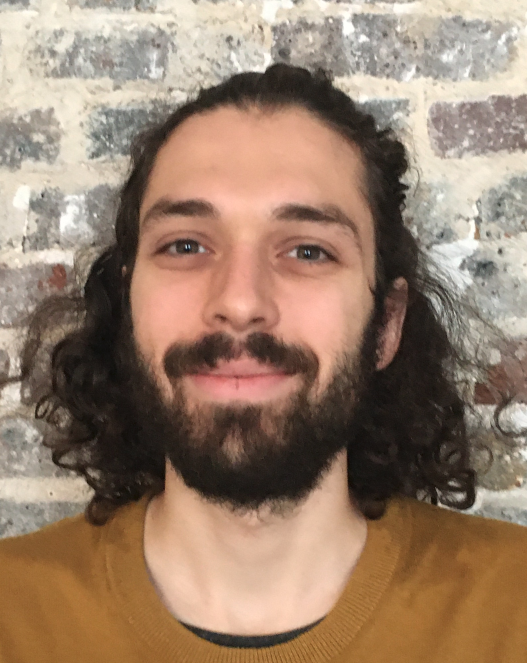}}{\small\quad {\bf Paul Marillonnet}
is a PhD student at Télécom SudParis, Institut Polytechnique de Paris, France.
He performs his research within the Entr'ouvert team, a cooperative specialized in providing Identity- and User-relationship-management free software to territorial collectivities and public administrations in France and in Belgium.
His field of research is the management of Personally-Identifiable Information (PII) by users of such collectivities and administrations, and the issues that arise when this shift in information governance happens.}\\[1mm]}

\noindent\parbox{8.3cm}{\parpic{\includegraphics[scale=.16]{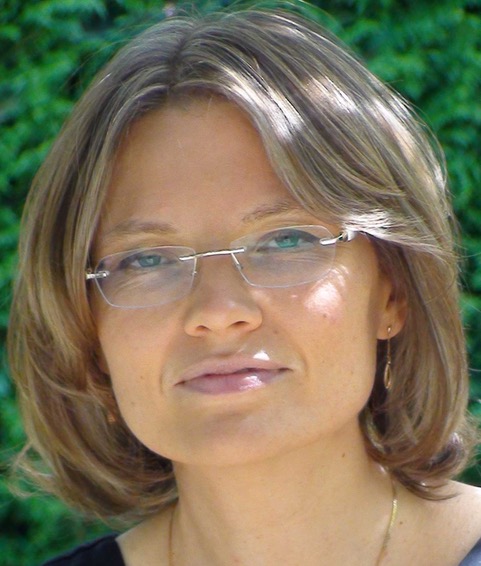}}{\small\quad {\bf Maryline Laurent}
works as a Professor at T\'el\'ecom SudParis, Institut Polytechnique de Paris, France. She leads the RST department (Telecommunication Networks and services) of T\'el\'ecom SudParis, and she is co-founder of the chair of Institut Mines-T\'el\'ecom on ``Values and Policies of Personal Information''.
She is area editor of Annals of Telecommunication journal, and editor of several books including ``Digital Identity Management'' (2015).
She lastly chaired IFIP WISTP 2019 conference on Information Security Theory and Practice.
Her research topics are related to network security and privacy applied to clouds, Internet of Things and identity management.}\\[1mm]}

\noindent\parbox{8.3cm}{\parpic{\includegraphics[scale=.07]{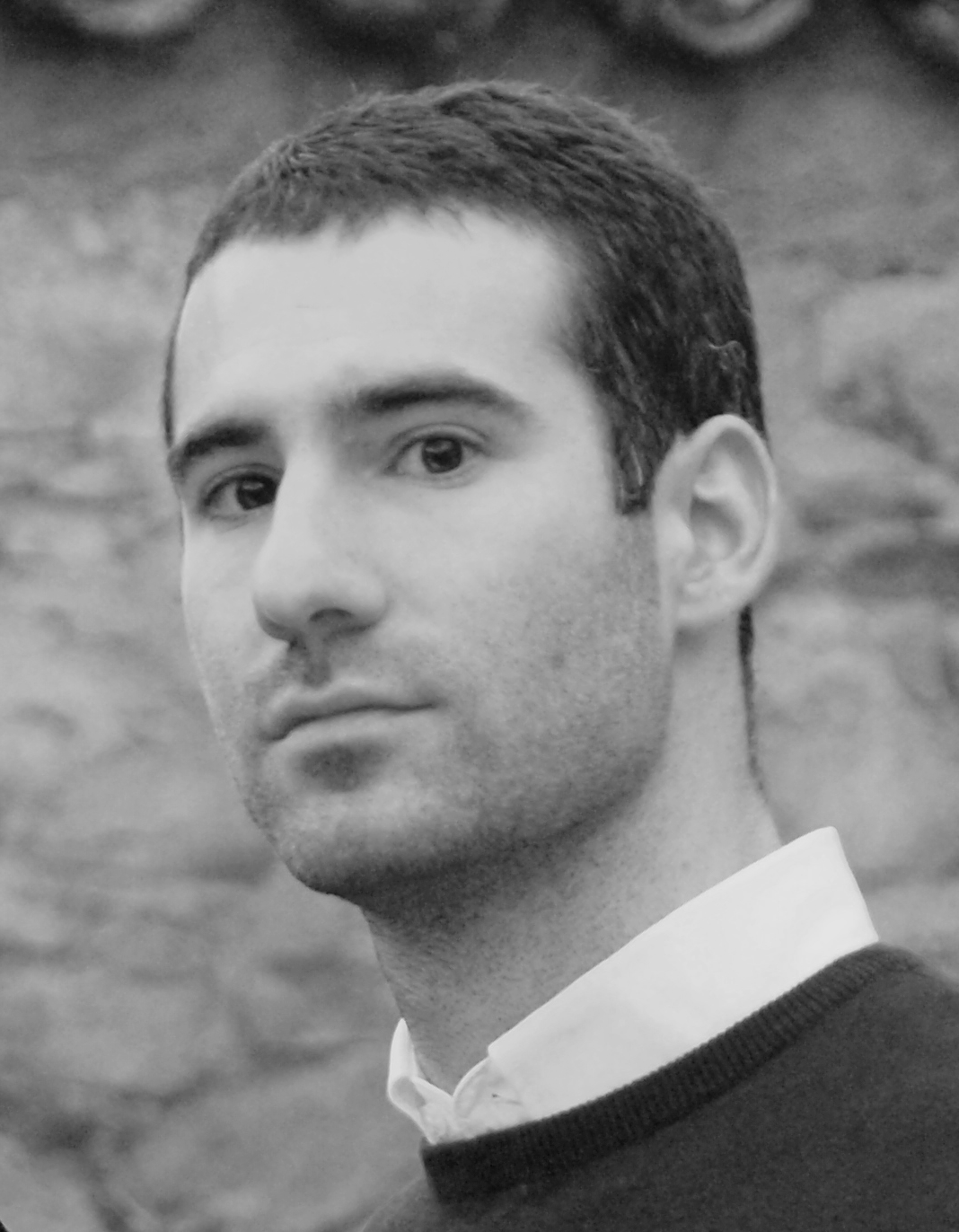}}{\small\quad {\bf Mikaël Ates}
is a project manager and research engineer, PhD in computer sciences, author of a thesis on digital identities and author of scientific articles in the fields of identity and personal data management, privacy and security.
He is coordinating both customer and research projects at Entr'ouvert and is contributing to these projects as an expert in digital-identity and user-relationship management.}\\[1mm]}

\label{last-page}

\end{multicols}
\end{document}